\documentclass[final,5p,twocolumn,10pt]{elsarticle} 

\usepackage{lineno}
\usepackage[draft]{hyperref}
\usepackage{enumitem}
\modulolinenumbers[5]

\usepackage{amsmath}
\usepackage{amssymb}
\usepackage{stmaryrd}
\usepackage{amsthm}
\usepackage{amsfonts}
\usepackage{dsfont}
\usepackage{graphicx}
\usepackage{upgreek}
\usepackage{bm}
\usepackage{color}
\usepackage{float}
\usepackage{tikz-cd}
\usepackage{relsize}

\usepackage{mathtools}
\usepackage{algorithm,algpseudocode} 
\usepackage{todonotes}

\usepackage{subcaption}

\usepackage{tabu}
\usepackage{dblfloatfix} 

\biboptions{numbers,sort&compress}

\newcommand{\bx}{{\mathbf{x}}}

\newcommand{\bt}{{\mathbf{t}}}
\newcommand{\bk}{{\mathbf{k}}}
\newcommand{\bu}{{\mathbf{u}}}

\newcommand{\bK}{{\mathbf{K}}}
\newcommand{\bff}{{\mathbf{f}}}

\newcommand{\indic}{{\mathbf{1}}}

\newcommand{\vol}{\mathsf{vol}}

\newcommand{\R}{{\mathds{R}}}

\newcommand{\SO}[1]{{\mathrm{SO}(#1)}}

\newcommand{\conf}{{\mathsf{C}}}

\newcommand{\targ}{\text{target}}
\newcommand{\acc}{\text{acc}}
\newcommand{\con}{\text{con}}
\newcommand{\unc}{\text{unc}}

\theoremstyle{definition}

\newcommand{\eq}[1]{(\ref{#1})} 
\newcommand{\com}[1]{} 

\journal{Computer-Aided Design}

\begin{document}

\begin{frontmatter}

\title{Topology Optimization for Manufacturing with Accessible Support Structures}

\author{ Amir M. Mirzendehdel, Morad Behandish, and Saigopal Nelaturi}
\address{\rm Palo Alto Research Center (PARC), 3333 Coyote Hill Road, Palo Alto, California 94304  \vspace{-15pt}}

\begin{abstract}
Metal additive manufacturing (AM) processes often fabricate a \textit{near-net} shape that includes the as-designed part as well as the sacrificial support structures that need to be machined away by subtractive manufacturing (SM), for instance multi-axis machining. Thus, although AM is capable of generating highly complex parts, the limitations of SM due to possible collision between the milling tool and the workpiece can render an optimized part non-manufacturable. We present a systematic approach to topology optimization (TO) of parts for AM followed by SM to ensure removability of support structures, while optimizing the part's performance. A central idea is to express the producibility of the part from the near-net shape in terms of {\it accessibility} of every support structure point using a given set of cutting tool assemblies and fixturing orientations. Our approach does not impose any artificial constraints on geometric complexity of the part, support structures, machining tools, and fixturing devices. We extend the notion of \textit{inaccessibility measure field} (IMF) to support structures to identify the inaccessible points and capture their contributions to non-manufacturability by a continuous spatial field. IMF is then augmented to the sensitivity field to guide the TO towards a manufacturable design. The approach enables efficient and effective design space exploration by finding nontrivial complex designs whose near-net shape can be 3D printed and post-processed for support removal by machining with a custom set of tools and fixtures. We demonstrate the efficacy of our approach on nontrivial examples in 2D and 3D.
\end{abstract}


\end{frontmatter}

\linenumbers

\section{Introduction} \label{sec_intro}

Many additive manufacturing (AM) technologies require sacrificial support structures for various reasons such as carrying unsupported weight before solidification and prevent ``drooling'' (e.g., fused deposition modeling), restraining the part to a substrate (e.g.,
stereo-lithography), or disposing the excess heat generated by intensive laser power to prevent residual stresses, thermal distortion, burning, and warping in metal AM (e.g., powder-bed fusion) \cite{jiang2018support}. Support structures can increase the AM cost directly and indirectly, especially in metal-based processes where material, fabrication time and energy, and post-fabrication cleanup costs are significant. For instance, recent studies have shown that depending on the heat treatment and porosity of support structures, support removal by machining can cause significant tool wear \cite{tripathi2018milling}. Nonetheless, support removal via milling remains popular in post-processing of metal AM parts \cite{holler2019direct,hintze2020finish}.

\begin{figure*} [t!]
   		\centering
   		\includegraphics[width=\linewidth]{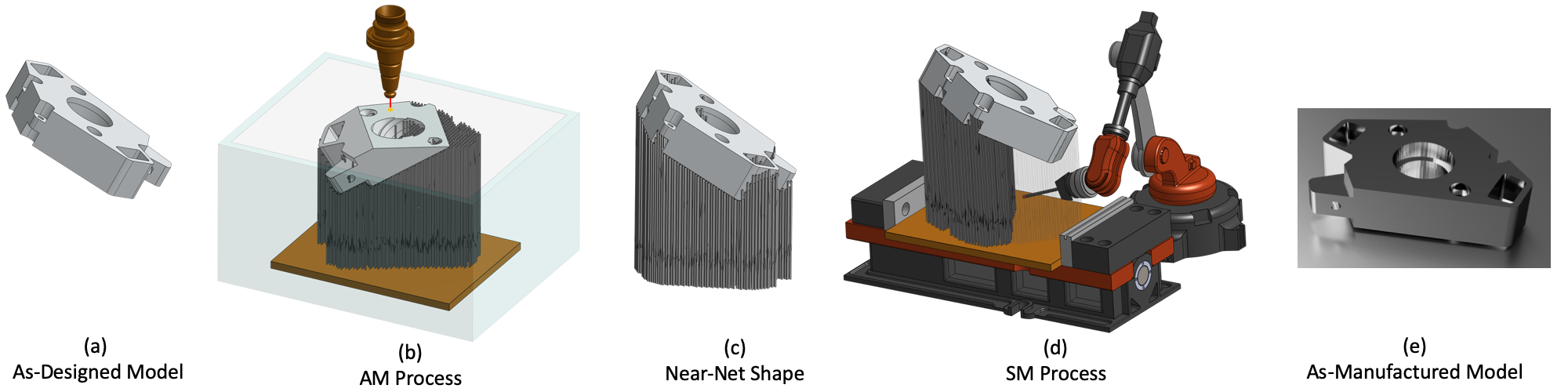}
   		\caption{Metal AM is typically a hybrid AM-then-SM process. The figure shows (a) an as-designed model; (b) an intermediate step in the AM process (e.g., powder-bed fusion \cite{bhavar2017review}) at a sub-optimal build orientation (selected arbitrarily for illustration); (c) the near-net including both the target part and support structured; (d) an intermediate step in the SM process (e.g., multi-axis machining with a vise fixture holding the build platform) to remove the support structures; and (e) the final as-manufactured part.} \label{fig_suppMachining}
   \end{figure*}

Despite many efforts in eliminating the support structures by optimizing the build orientation or the part, in many practical cases it is not feasible to fabricate AM parts without support structures. In metal AM, the sacrificial supports need to be removed by subtractive manufacturing (SM) such as multi-axis computer numerically-controlled (CNC) machining. This common ``hybrid'' (AM-then-SM) workflow is illustrated in Fig. \ref{fig_suppMachining}. 

Topology optimization (TO) is widely used as an automated design tool to search the expanded design space of AM parts for high-performance and light-weight designs. TO commonly generates parts of complex geometry with intricate features that can pose manufacturing challenges.
The novel concept of inaccessibility measure field (IMF) was introduced in \cite{mirzendehdel2019exploring,mirzendehdel2020topology} to locally quantify one's ability to access different points of an intermediate design using a set of cutting tool assemblies and fixturing orientations. The continuous IMF was used to penalize the TO sensitivity field, steering the updates to improve not only physics-based performance criteria (e.g., stiffness), but also manufacturability (e.g., machining).
In this paper, we extend this TO framework from a pure-SM to an AM-then-SM workflow in which the optimization process is constrained to generate designs whose corresponding near-net shape is accessible at every support structure point for a given set of cutting tool assemblies and fixturing orientations. The near-net shape is defined as the super-set of the final as-manufactured shape and its corresponding support structures. Our approach applies to parts, support structures, tool assemblies, fixture, and build platforms of arbitrary shape, as well as high-axis machining degrees of freedom.

\subsection{Related work} \label{sec_lit}
In this section, we review recent advances in design for manufacturing (DfM), particularly design for AM (DfAM), and hybrid (additive followed by subtractive) manufacturing (DfHM) with a focus on TO as an automated design tool. 

Over the past few years, TO has become increasingly popular in industrial engineering applications to recommend viable topologies in early stages of part design \cite{liu2018current,dbouk2017review}. The main challenge in adoption of TO for detailed/final design is to fill the gap between the `as-designed' and `as-manufactured' models that results in a costly trial-and-error cycle. 

Considering both traditional and modern manufacturing processes, numerous DfM and DfAM approaches have been developed, where typically the manufacturing constraints are simplified under reasonable assumptions, reduced to straight-forward design rules, and are imposed through filtering/projection methods \cite{norato2015geometry}. Traditional manufacturing constraints include ensuring minimum feature size \cite{sigmund2007morphology,guest2009imposing,zhou2015minimum}, casting \cite{harzheim2006review,li2018topology}, laser-cutting/extrusion \cite{ishii2004topology,hoang2020extruded}, and milling \cite{langelaar2019topology,gaynor2020eliminating,morris2020subtractive}.  
For instance, proposed approaches for incorporating multi-axis machining constraint in TO typically assume simple geometries for machine tools, ignore surrounding fixturing devices, and abstract tool motion by tracing straight lines. Recently, a general approach was presented \cite{mirzendehdel2019exploring,mirzendehdel2020topology} to incorporate accessibility constraints for high-axis machining via continuous measures, using well-established configuration space ($\conf-$space) concepts widely used in spatial planning and robotics \cite{lozano1990spatial}, with no restrictive assumptions on geometric complexity (of part and tool), tool motion, or machining setup. On the other hand, DfAM methods have mainly focused on reducing support material by optimizing the part geometry \cite{mirzendehdel2016support,langelaar2016topology,wu2016self,qian2017undercut,allaire2017structural}, optimizing support structures \cite{malekipour2018heat,zhou2020topology,wang2020optimizing}, optimizing build orientation \cite{chandrasekhar2020build,wang2020simultaneous}, TO for anisotropic materials \cite{mirzendehdel2018strength,zhang2017role}, minimizing residual stresses \cite{cheng2019utilizing,allaire2018taking,miki2020topology}, and geometric post-processing \cite{swierstra2020automated,subedi2020review}.

DfHM approaches are still fairly limited and focus on AM-then-SM processes. For example, Liu et al. \cite{liu2017topology} proposed a TO method by imposing a casting constraint on free-form AM design domain and feature fitting on shape-preserved segments for machined surfaces. Lou et al. \cite{luo2020additive} presented a TO formulation based on a nonlinear virtual temperature method to ensure that inaccessible enclosed voids are self-supporting, while allowing the free boundary to have overhang surfaces. No accessibility analysis with respect to tool geometry or machining setup was considered.
More detailed reviews of manufacturing-oriented TO can be found in \cite{liu2016survey,liu2018current,plocher2019review}. 

\subsection{Contributions \& Outline} 

This paper proposes a TO method for hybrid AM-then-SM manufacturing, where given a prescribed build orientation and multi-axis machining setup, the near-net shape corresponding to the optimized design is guaranteed to have removable supports. The multi-axis machining setup comprises a collection of machining tool assemblies, fixturing devices, and motions including translations and rotations. There is no artificial and restricting assumptions on geometric complexity of tools and fixtures, or motion. Our proposed manufacturability-constrained optimization incorporates a sophisticated manufacturing constraint at early stages of design to drastically reduce downstream trial and error costs. 
More specifically, the contributions of this paper are:

\begin{enumerate}
	\item Extending the `inaccessibility measure field' to AM-then-SM hybrid manufacturing to capture accessibility of support structures.
	\item Formulating a density-based TO framework that incorporates accessibility constraint for support removal given a realistic multi-axis machining setup  with multiple geometrically complex cutting tool assemblies, fixturing devices, and arbitrary motions.
	\item Proposing a support removal planning algorithm to gradually remove supports until the part is detached from the platform.
	\item Demonstrating the effectiveness of our method by solving multiple benchmark and realistic examples in 2D and 3D.
\end{enumerate}

\section{Proposed Method} \label{sec_method}
In this section, we will extend the TO formulation, specifically density-based TO, to incorporate accessibility of support structures given a multi-axis machining setup. 

\subsection{Topology Optimization Formulation}

The design problem can be formulated as the following constrained optimization problem:

\begin{subequations} \label{eq_TOproblemAcc}
	\begin{align}
	\mathop{\text{Minimize}}\limits_{\Omega \subseteq \Omega_0} \quad &
	\varphi(\Omega), \label{eq_TOproblemAcc_a} \\
	\text{such that} \quad 
	& [\textbf{K}_\Omega][\textbf{u}_\Omega]=[\bff], \label{eq_TOproblemAcc_b} \\
	& V_\Omega \leq V_{\targ}, \label{eq_TOproblemAcc_c}\\ 
	& V_{\Gamma}/V_S \le \epsilon, \label{eq_TOproblemAcc_d} 
	\end{align}
\end{subequations}
where the following definitions and notations are used:
\begin{itemize}
	\item [] $\Omega_0 \subset \R^3$:\quad 3D design domain;
	\item [] $\Omega \subseteq \Omega_0$:\quad 3D pointset representing the (intermediate or final) design, to be updated iteratively;
	\item [] $\rho_\Omega: \Omega_0 \to [0, 1]$:\quad The density field that implicitly represents the design as a superlevel-set;
	\item [] $\phi(\Omega) \in \R$:\quad The objective function value (e.g., compliance) for a given design;
	\item [] $[\bK_\Omega] \in \R^{n \times n}$:\quad Finite element stiffness matrix for a voxelization of the domain into $n$ elements;
	\item [] $[\bu_\Omega] \in \R^n$:\quad Displacement vector for the voxelization;
	\item [] $[\bff]  \in \R^n$:\quad External force vector for the voxelization;
	\item [] $V_\text{target} \in \R$: Target volume upperbound;
	\item [] $V_\Omega \in \R$: The current design's volume;
	\item [] $O \subset \R^3$: Obstacle geometry defined as the union of part, platform, and fixturing devices ($O = \Omega~\cup~P~\cup~F$), where there must be no collision between cutting tools and obstacle $O$;
	\item [] $V_S \in \R$: \quad Support structure volume;
	\item [] $V_\Gamma \in \R$:\quad Secluded (inaccessible) support volume; and
	\item [] $\epsilon > 0$:\quad A small tolerance for non-manufacturability (e.g., to enable  convergence in the presence of numerical inaccuracies).
\end{itemize}
In Section \ref{sec_acc}, we explain the accessibility analysis required to find the secluded supports $\Gamma$ in terms of the cutting tool assemblies $T$, obstacle $O$, and support structures $S$.

Without considering the support accessibility constraint in \eq{eq_TOproblemAcc_d}, the constrained TO problem defined in \eq{eq_TOproblemAcc_a} through \eq{eq_TOproblemAcc_c} can be expressed as the following Lagrangian minimization problem: 
\begin{equation}
	\mathcal{L}_{\Omega}:= \varphi(\Omega)
	+ ~\lambda_1 (\dfrac{V_\Omega}{V_{\targ}} - 1)
	+ [\lambda_2]^\mathrm{T} \Big([\textbf{K}_\Omega][\bu_\Omega] - [\bff]\Big).
	\label{eq_Lag}
\end{equation}
Taking the derivative w.r.t. design variables denoted by the prime symbol $(\cdot)'$ and using the chain rule we have:
\begin{align}
	\mathcal{L}_\Omega' &= \varphi'(\Omega) + \lambda_1
	\dfrac{{V}_\Omega'}{V_{\targ}} + [\lambda_2]^\mathrm{T}
	\Big([\textbf{K}_\Omega][\bu_\Omega]\Big)',\\
	& = \Big([ \dfrac{\partial \varphi}{\partial \bu} ]+ [\lambda_2]^\mathrm{T}
	[\textbf{K}_\Omega]\Big)[\bu_\Omega'] \nonumber \\
	& +\lambda_1 \dfrac{{V}_\Omega'}{V_{\targ}} + [\lambda_2]^\mathrm{T}
	[\textbf{K}_\Omega'][\bu_\Omega]. \label{eq_chain_rule}
\end{align}
\eq{eq_chain_rule} reduces to \eq{eq_Lag_prime} via the adjoint method
\cite{Bendsoe2009topology}:
\begin{align}
	&\mathcal{L}_\Omega' = \lambda_1 \dfrac{{V}_\Omega'}{V_{\targ}}  +
	[\lambda_2]^\mathrm{T} [\textbf{K}_\Omega'][\bu_\Omega], \label{eq_Lag_prime}\\
	&\text{if} ~ [\lambda_2] := -[\textbf{K}_\Omega]^{-1}[\dfrac{\partial
		\varphi}{\partial \bu}]. \nonumber
\end{align}
Here the objective function is compliance, i.e., $\varphi(\Omega) := [\bff]^\mathrm{T}[\bu_\Omega]$, thus
$[\lambda_2] = -[\bu_\Omega]$.

\subsection{Accessibility Analysis for Support Structures} \label{sec_acc}

In this section, we briefly discuss the IMF as a means to quantify collision at each point $\bx \in \Omega_0$ given a design $\Omega$, a set of tool assemblies $T \subset \R^3$, sampled cutting points $K \subset \R^3$ and rotations $ \Theta \subset \SO{3}$, platform $P 
\subset \R^3$, and fixturing devices $F \subset \R^3$. In particular, we review the convolution-based formulation of IMF, more details can be found in \cite{mirzendehdel2020topology}.

We define the collision between obstacle $O=\Omega \cup P \cup F$ and cutting tool $T$ under rotation $R \in \Theta$ and translation $\bt$ as:
\begin{equation}
\vol\big[O \cap (R, \bt)T \big] = \langle \indic_O, \indic_{(R, \bt)T} \rangle
= (\indic_{O} \ast \tilde{\indic}_{RT}) (\bt), \label{eq_ident}
\end{equation}
where $\vol[\cdot]$ denotes the volume of a 3D pointset which can be expressed as the inner product of the indicator functions of the obstacle $\indic_{O}$ and the transformed (rotated and translated) tool $\indic_{(R, \bt)T}$. It has been shown in \cite{lysenko2010group} that the inner product can be written as a convolution for objects in relative translation, namely, the obstacle $O$ and the rotated tool $RT$. The convolution operator $\ast$ is defined as:

\begin{equation}
(\indic_{O} \ast \tilde{\indic}_{RT}) (\bt) = 
\int_{O}\indic_{O}(\bx)\indic_{-RT}(\bt-\bx)dv[\bx], \label{eq_conv}
\end{equation}
where $\indic_{-RT}(\bt) = \indic_{RT}(-\bt) = \tilde{\indic}_{RT}(\bt)$ is the reflection of the rotated tool about the origin, hence $\indic_{-RT}(\bt-\bx) = \indic_{RT}(\bx-\bt) = \tilde{\indic}_{RT}(\bt-\bx)$ is the indicator function of the moving object (i.e., rotated tool assembly) translated to the query points $\bx \in O$. The convolution of \eq{eq_conv} can be efficiently evaluated via Fast Fourier Transforms (FFT).

\begin{figure}[ht!]
	\centering
	\includegraphics[width=\linewidth]{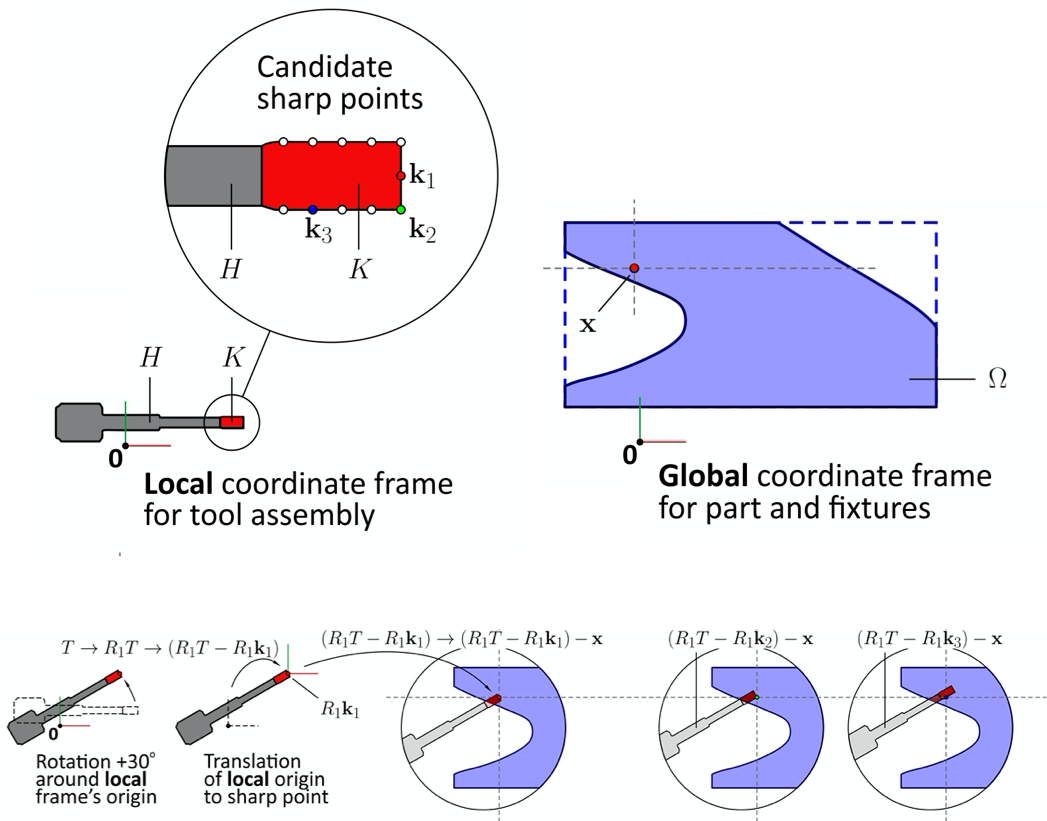}
	\caption{Consider a 2D part and a 2D tool assembly. At a given query
		point $\bx \in \Omega_0$, the IMF is computed by looking at different rotations, e.g., 
		$R_1 \in \SO{2}$ of the tool assembly. For the oriented tool $T \to R_1 T$, the origin is shifted to
		different sharp points $\bk_1, \bk_2, \bk_3, \ldots  \in K$; e.g., for the
		(rotated) sharp point $R_1 \bk_j \in R_1 K$, the tool is translated as $R_1 T
		\to (R_1T - R_1 \bk_j)$. The new origin is brought in contact with
		the query point, hence $(R_1T - R_1 \bk_j) \to (R_1T - R_1 \bk_j) -\bx$. This is repeated for all candidate sharp points and the IMF is computed as the minimum over all rotations and sharp points.}
	\label{fig_sharpPts}
\end{figure}

As illustrated in Fig. \ref{fig_sharpPts}, we define the IMF as a mapping $f_\text{IMF}: \R^3 \to \R$ that assigns a scalar value to each point $\bx \in \R^3$, where for a given tool assembly $T = (H \cup K)$ at orientation $R$ we take a pointwise minimum of shifted convolutions for different choices of sharp points in the cutter $\bk \in K$:
\begin{equation}
f_\text{IMF}(\bx; O, R, T, K) :=  \min_{\bk \in K} ~ {\vol\big[ O \cap (R, \bx) (T - \bk) \big]}
. \label{eq_imf_RotToolIndic}
\end{equation}
Since, the maximum collision volume happens when the entire tool assembly $T$ collides with the obstacle $O$, $f_\text{IMF}$ defined in \eq{eq_imf_RotToolIndic} is normalized so that its value is within [0,1] everywhere. The normalized IMF value denoted as $\bar{f}_\text{IMF}$ can be written as:
\begin{equation}
	\bar{f}_\text{IMF}(\bx; O, R, T, K) :=  \min_{\bk \in K} ~ \dfrac{\vol\big[ O \cap (R, \bx) (T - \bk) \big]}{\vol\big[T\big]}
	. \label{eq_imf_RotToolIndicNormalized}
\end{equation}

To use \eq{eq_imf_RotToolIndicNormalized} in a density-based TO loop, we define the function $\rho^{}_O: O \to [0, 1]$ as:
$\rho^{}_O(\bx) := \rho^{}_\Omega(\bx) + \indic_P(\bx) + \indic_F(\bx)$, in which
$\rho^{}_\Omega(\bx)$ is obtained directly from TO. \footnote{Also, note that the implicit assumption here is that the fixture does not collide with neither the platform nor the part -- otherwise the function will not be within [0,1].}  \footnote{The + sign indicates disjoint union in the limit when $\rho_\Omega = 1_\Omega, \text{ and } 1_\Omega + 1_P + 1_F = 1_{(\Omega \cup P \cup F)}$.}

The IMF in terms of intermediate densities is defined as:  
\begin{equation}
\bar{f}_\text{IMF}(\bx; \rho^{}_O, T, K) := \min_{\bk \in K}
~ \dfrac{(\rho^{}_{O} \ast \tilde{\indic}_{RT})(\bx - R\bk)} {\vol\big[T\big]}. \label{eq_imf_RotTool}
\end{equation}
Since $\rho^{}_O \in [0,1]$, $0 \le \bar{f}_\text{IMF}(\bx,\rho^{}_O) \le \bar{f}_\text{IMF}(\bx; O) \le 1$. It is also worth noting that the \eq{eq_ident} relates ``volume'' of collision to a convolution and is only valid for indicator functions. Thus, the convolution of \eq{eq_imf_RotTool} is not exactly a volumetric measure, however it is bounded by \eq{eq_imf_RotToolIndicNormalized} and as the TO approaches a solid-void design, the values of the two equations converge. The procedure for computing IMF for a rotated tool is described in Algorithm \ref{alg_imf1}.  

\begin{algorithm} [ht!]
	\caption{Compute $[\bar{f}^{}_{\text{IMF}}]$ for a rotated tool.}
	\begin{algorithmic}
		\Procedure{IMF}{$[\rho^{}_{O}],R,[\indic_{T}],[\indic_{K}]$}
		\State $[\indic_{RT}] \gets \Call{Rotate}{[\indic_{T}], R}$
		\Comment{Re-sampling}
		\State $[\tilde{\indic}_{RT}] \gets \Call{Reflect}{[\indic_{RT}]}$
		\State $[g] \gets \Call{Convolve}{[\rho^{}_O], [\tilde{\indic}_{R T}]}$
		\Comment{FFT-based}
		\State Initialize $[{f}_\text{IMF}] \gets [0]$
		\ForAll{$\bk \in K$}
		\State $[h] \gets \Call{Translate}{[g], -R \bk}$
		\State $[{f}_\text{IMF}] \gets \min([f_\text{IMF}], [h])$
		\Comment{Over sharp points}
		\EndFor
		\State $[\bar{f}_\text{IMF}] \gets [{f}_\text{IMF}]/ \vol[T]$
		\State\Return{($[\bar{f}_\text{IMF}]$)}
		\EndProcedure 
	\end{algorithmic} \label{alg_imf1}
\end{algorithm}

Next, given available orientations $\Theta \subset \SO{3}$ for each tool assembly, $f_\text{IMF}$ for each tool can be written as:
\begin{equation}
\bar{f}_\text{IMF}(\bx; \rho^{}_O) := \min_{R \in \Theta} \bar{f}_\text{IMF}(\bx; \rho^{}_O, T, K) \label{eq_imf_Tool}
\end{equation}

Subsequently, the combined IMF for all
$n_T$ tool assemblies is computed as:
\begin{equation}
\bar{f}_\text{IMF}(\bx; \rho^{}_O) := \min_{1 \leq i \leq n_T}
\bar{f}_\text{IMF}(\bx; \rho^{}_O, T_i, K_i) \label{eq_imf_Overall}
\end{equation}
IMF of \eq{eq_imf_Overall} can be computed as described in Algorithm \ref{alg_imf2}.  

\begin{algorithm} [ht!]
	\caption{Compute overall $[\bar{f}^{}_{\text{IMF}}]$ of \eq{eq_imf_Overall}.}
	\begin{algorithmic}
		\Procedure{IMF}{$[\rho^{}_{\Omega}],[\indic_P], [\indic_F], [\indic_{H_i}],
			[\indic_{K_i}], \{\Theta_i\}; n_T$}
		\State Define $[\rho^{}_{O}] \gets [\rho^{}_{\Omega}] + [\indic_P]+[\indic_{F}]$
		\Comment{Disjoint union}
		\State Initialize $[\bar{f}_\text{IMF}] \gets [0]$ \Comment{IMF for all the tools}
		\For{$i \gets 1$ to $n_T$}
		\State Define $[\indic_{T_i}] \gets [\indic_{H_i}] + [\indic_{K_i}]$
		\Comment{Implicit union}
		\ForAll{$R \in \Theta_i$}
		\State $\gamma \gets \Call{IMF}{[\rho^{}_{O}],R,[\indic_{T_i}],[\indic_{K_i}]}$ \Comment{Alg. \ref{alg_imf1}}
		\State $[\bar{f}_\text{IMF}] \gets \min([\bar{f}_\text{IMF}],\gamma) $ \Comment{ \eq{eq_imf_Overall}} 
		\EndFor	
		
		\EndFor
		\State\Return{($[\bar{f}_\text{IMF}]$)}
		\EndProcedure 
	\end{algorithmic} \label{alg_imf2}
\end{algorithm}

\subsection{Topology Optimization for Accessible Supports}

To incorporate the support accessibility constraint for multi-axis machining, we modify
the sensitivity field $	\mathcal{S}_\Omega$ as follows:
\begin{equation}
\mathcal{S}_\Omega := (1-w_{\acc}) ~ \bar{\mathcal{S}}_\varphi + w_{\acc} ~
\bar{\mathcal{S}}_{\text{IMF}}, \label{eq_filtered_sens}
\end{equation}
where $0 \le w_{\acc} < 1$ is the filtering weight for accessibility. In this paper, $w_{acc}$ gradually increases from 0 to 0.5. This allows the TO to start from a close-to-optimal design w.r.t. performance and gradually modify it w.r.t. support accessibility.
$\bar{\mathcal{S}}_\varphi$ is the normalized sensitivity field with respect to the
objective function, 
noting that the volume constraint is satisfied with the optimality criteria iteration \cite{sigmund200199}. $\bar{\mathcal{S}}_{\text{IMF}}$ is the
normalized accessibility filter defined in terms of the normalized IMF over the design obtained with the physical density $\tilde{\rho}_\Omega$. The physical density $ \tilde{\rho}_\Omega$ is obtained from the Heaviside projection \cite{Andreassen2011efficient}:
\begin{equation}
	\tilde{\rho}^{}_\Omega (\bx) = 1 - e^{-\beta \rho^{}_\Omega(\bx)}+ \rho{}_\Omega(\bx)
	e^{-\beta}, \label{eq_Heaviside}
\end{equation}
 Our initial TO experiments showed that a layer-wise penalization of the support inaccessibility measure field helps the optimizer to converge to a better solution. The heuristic essentially gives preference in retaining material close to platform if performance sensitivity and collision values are comparable. Further, since the constraint is imposed as a soft constraint, in some pathological cases a small amount of inaccessible supports might remain towards the end of optimization, where there may be low-density regions with a small non-zero IMF value, for which the sensitivity filter is not enough to make the barely-secluded regions fully accessible.
In this case, we continue optimization while also imposing a filter directly on density values by gradually increasing the density at the secluded regions. In our experiments, this helps TO to converge to a valid solution within a few iterations.
\begin{equation}
\bar{\mathcal{S}}_{\text{IMF}} (\bx) :=
\begin{dcases}
-\bar{L}^q f_\text{IMF}(\bx; \tilde{\rho}^{}_O) &  \text{if} ~ \bx \in \Omega \cup S , \\
0 & \text{otherwise}.
\end{dcases} \label{eq_accFilter}
\end{equation}	

\noindent where for the $k^{th}$ layer of $n_z$ layers, $\bar{L}_k^q = \left(\dfrac{n_z-k+1}{n_z}\right)^q$. For all examples, we used a layer-wise penalization coefficient of $q=4$. 
We use $\beta := 1$ for 2D and 3D examples of Section
\ref{sec_results}, which in our implementation seems to provide a good physical density with sufficient smoothness for computing IMF.

Algorithm \ref{alg_TO} describes the TO with support accessibility constraint, where $\textbf{b}$  and $\alpha$ denote the build direction and the overhang angle for support generation in AM.
\begin{algorithm} [ht!]
	\caption{TO with multi-axis machining constraint for support removal.}
	\begin{algorithmic}
		\Procedure{TO}{$\Omega_0,V_{target},\textbf{b},\alpha,q$}
		\State Initialize $[\rho_\Omega] \gets [V_\targ /
		\Call{Integral}{[\indic_{\Omega_0}]}]$
		\State Initialize $\Delta \gets \infty$
		\State Initialize $iter \gets 0$
		\State Initialize $w_{acc} \gets 0$
		\State ${S_0} \gets \Call{\text{SuppGen}}{\Omega_0,\textbf{b},\alpha}$
		\Comment{Support gen.}
		\State $[\indic_{N_0}] \gets [\indic_{\Omega_0}][ \indic_{S_0}]$ \Comment{Initial near-net shape}
		\State $\bar{L}^q \gets \Call{LayerCoefs}{[\indic_{N_0}]}$ \Comment{$\bar{L}_k = \dfrac{n_z-k+1}{n_z}$}
		\While {$\Delta > \delta$ \textbf{and} $iter < I$ }
		\State $[\tilde{\rho}^{}_\Omega] \gets \Call{Heaviside}{[\rho_\Omega], \beta}$
		\Comment{Projection}
		\State ${S} \gets \Call{\text{SuppGen}}{[\tilde{\rho}^{}_\Omega],b,\alpha}$
		\Comment{Support gen.}
		\State $[\indic_N] \gets [\indic_\Omega] [\indic_S]$ \Comment{Near-net shape}
		\State $[\bu] \gets \Call{\text{FEA}}{[\tilde{\rho}^{}_\Omega], [\bff]}$
		\Comment{Solve FEA}
		\State $\varphi(\Omega) \gets \Call{\text{Evaluate}}{[\rho^{}_\Omega],
			[\bu]}$ \Comment{Obj. func.}
		\State $[\bar{\mathcal{S}}_\varphi] \gets
		\Call{\text{Gradient}}{[\rho^{}_\Omega], [\bu]}$\Comment{Sensitivity}
		\State $[{f}_\text{IMF}] \gets \Call{IMF}{[\tilde{\rho}^{}_\Omega], [\indic_F], \ldots; n_T}$
		\State \Comment{Call Algorithm \ref{alg_imf2} with obvious arguments}
		\State $[\indic_\Gamma] \gets [\indic_S] [{f}_\text{IMF} > \lambda]$
		\State $[\bar{\mathcal{S}}_\text{IMF}] \gets [\indic_N] [-\bar{L}^q {f}_\text{IMF}]$
		\State $[\mathcal{S}] \gets (1-w_{\acc}) [\bar{\mathcal{S}}_\varphi]
		+ w_{\acc} [\bar{\mathcal{S}}_{\text{IMF}}]$
		\Comment{ \eq{eq_filtered_sens}}
		\State $[\rho_\Omega^{\text{new}}] \gets
		\Call{\text{Update}}{[\rho_\Omega], [\mathcal{S}]} $ 
		\Comment{Update density}
		\State $\Delta \gets \Call{Integrate}{[\rho_\Omega^{\text{new}}]
			- [\rho_\Omega]}$ \Comment{Vol. diff.}
		\State $iter \gets iter + 1$ \Comment{Iter.  counter}
		\State $[\rho_\Omega] \gets [\rho_\Omega^{\text{new}}]$	\Comment{For next iteration}
		\If{$iter > I_{acc}$}  \Comment{e.g., $I_{acc} = 20)$}
			\State $w_{acc} \gets w_{acc} + 0.01$ \Comment{Update weight}
		\EndIf
	\If{$iter > I_{\rho}$}  \Comment{e.g., $I_{\rho} = 150$}
		\State $\rho_{\Gamma}^{} = \min (\rho_{\Gamma}^{} + 0.5, 1.0)$\Comment{Penalize density at secluded regions}
		\State $\bar{\mathcal{S}}_{\Gamma} = \min (-\bar{L}^q, \bar{\mathcal{S}}_{\Gamma})$ 	\Comment{Penalize sensitivity at secluded regions}
	\EndIf
		\EndWhile
		\State\Return{$[\rho_\Omega]$}
		\EndProcedure 
	\end{algorithmic} \label{alg_TO}
\end{algorithm}
\section{Results} \label{sec_results}
In this section, we will present benchmark and realistic examples in 2D and 3D. All results are generated using a density-based implementation, where the optimality criteria method was used to update the density field. \\
All examples are run on a desktop machine with \textsf{Intel Corei7-7820X} CPU with 8 processors running at 4.5 GHz, 32 GB of host memory, and an \textsf{NVIDIA GeForce GTX 1080} GPU with 2,560 \textsf{CUDA} cores and 8 GB of device memory.
\subsection{Benchmark Example: Cantilever Beam}

\begin{figure}[ht!]
	\centering
	\includegraphics[width=\linewidth]{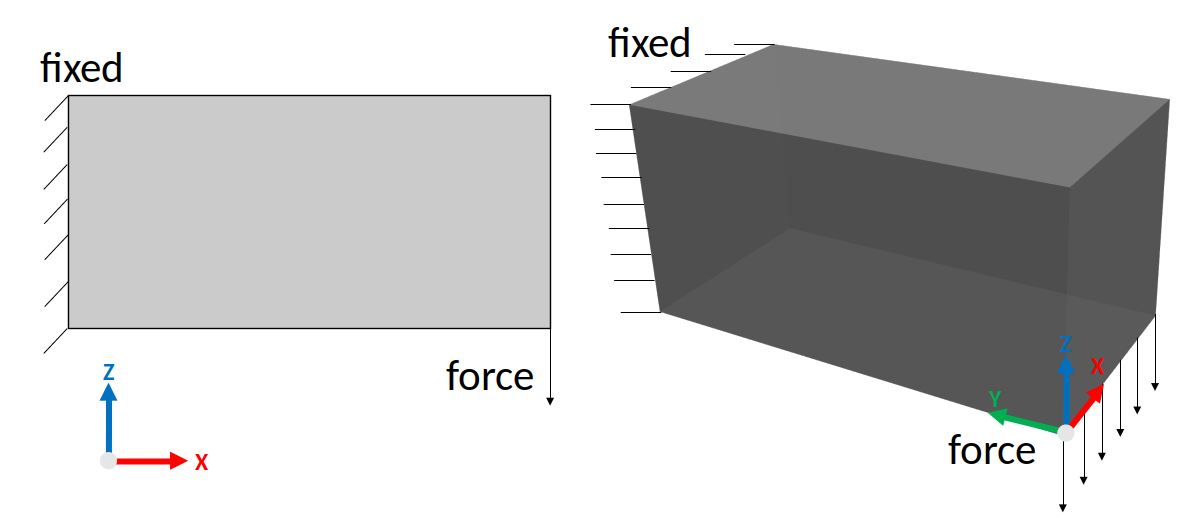}
	\caption{Cantilever beam boundary conditions in 2D and 3D.}
	\label{fig_beamLoading}
\end{figure}

\begin{figure}[ht!]
	\centering
	\includegraphics[width=\linewidth]{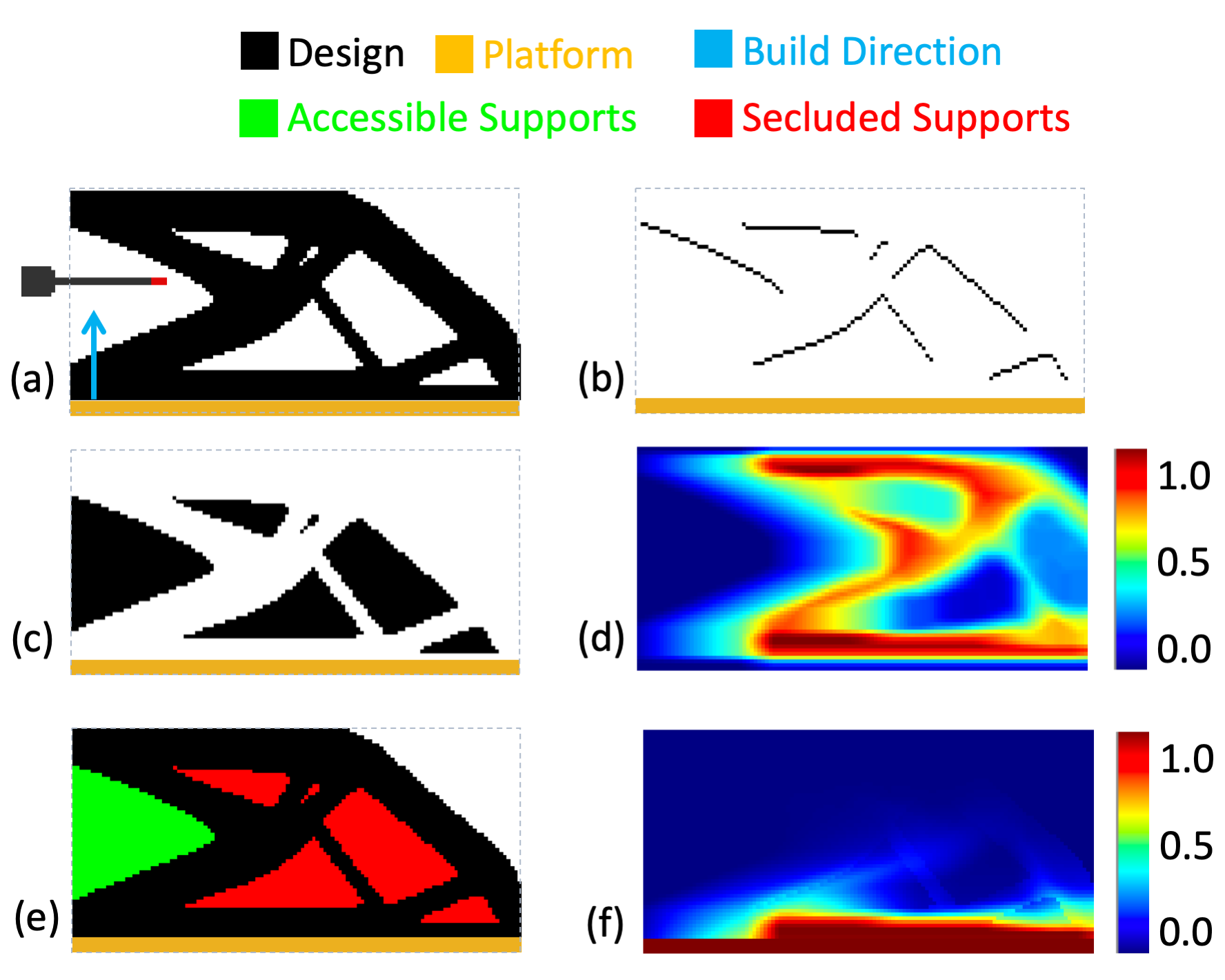}
	\caption{TO for cantilever beam in 2D \textit{without} accessibility constraint on support structures for $90^\circ$ overhang angle: (a) optimized cantilever beam at volume fraction of 0.5 and the oriented end-mill tool, (b) overhang points, (c) support structures (d) normalized IMF, (e) accessible and secluded supports, and (f) filtered IMF.}
	\label{fig_mfgAnalysis_beam2D}
\end{figure}

\begin{figure}[ht!]
	\centering
	\includegraphics[width=\linewidth]{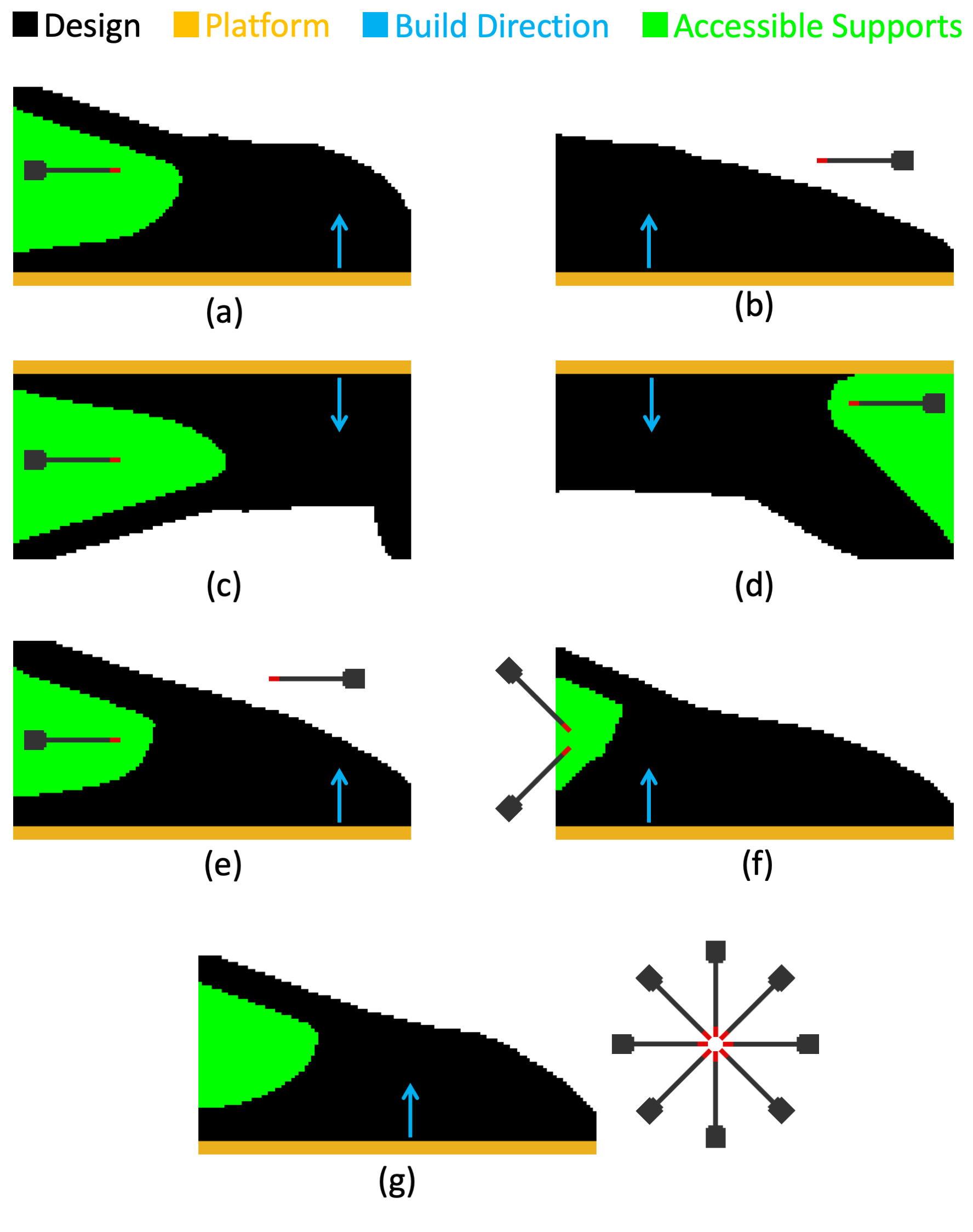}
	\caption{TO for cantilever beam in 2D \textit{with} accessibility constraint on support structures for $90^\circ$ overhang angle: (a) $0^\circ$ tool angle with $+Z$ build orientation, (b) $180^\circ$ tool angle with $+Z$ build orientation, (c) $0^\circ$ tool angle with $-Z$ build orientation, (d) $180^\circ$ tool angle with $-Z$ build orientation, (e) $\left\{ 0^\circ, 180^\circ \right\}$ tool angle with $+Z$ build orientation, (f) $\pm45^\circ$ tool angle with $+Z$ build orientation, (g) $\left\{ 0^\circ, 180^\circ, \pm 45^\circ, \pm 135^\circ, \pm 90^\circ \right\}$ tool angle with $+Z$ build orientation.}
	\label{fig_benchmark2D}
\end{figure}

\begin{figure}[ht!]
	\centering
	\includegraphics[width=\linewidth]{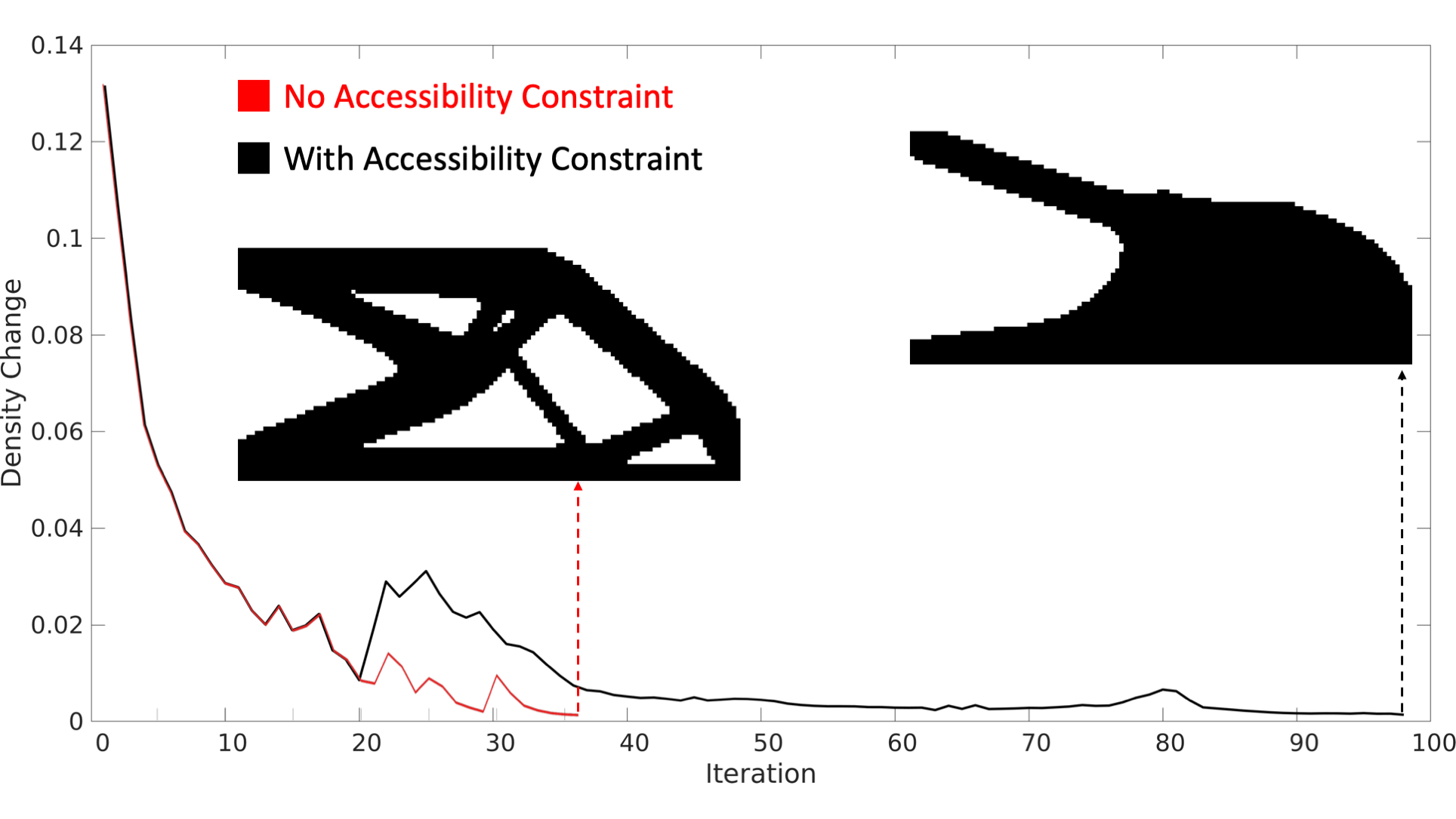}
	\caption{Density convergence plots for cantilever beam TO of Fig. \ref{fig_benchmark2D}a without and with support accessibility constraint for $90^\circ$ overhang angle.}
	\label{fig_densityConvergenceBeam2D}
\end{figure}

\begin{figure}[ht!]
	\centering
	\includegraphics[width=\linewidth]{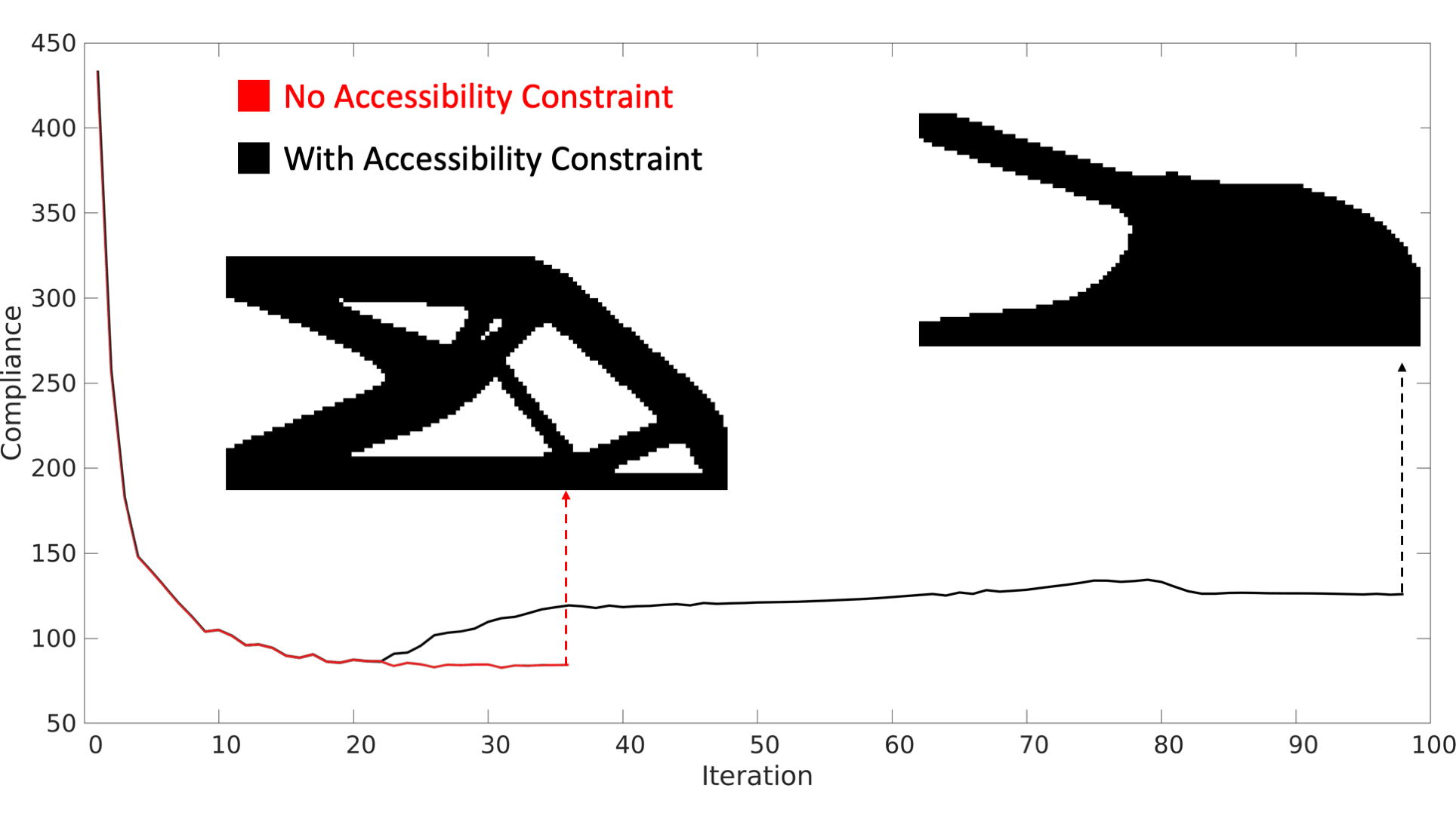}
	\caption{Compliance convergence plots for cantilever beam TO of Fig. \ref{fig_benchmark2D}a without and with support accessibility constraint. }
	\label{fig_complianceConvergenceBeam2D}
\end{figure}

\begin{figure}[ht!]
	\centering
	\includegraphics[width=\linewidth]{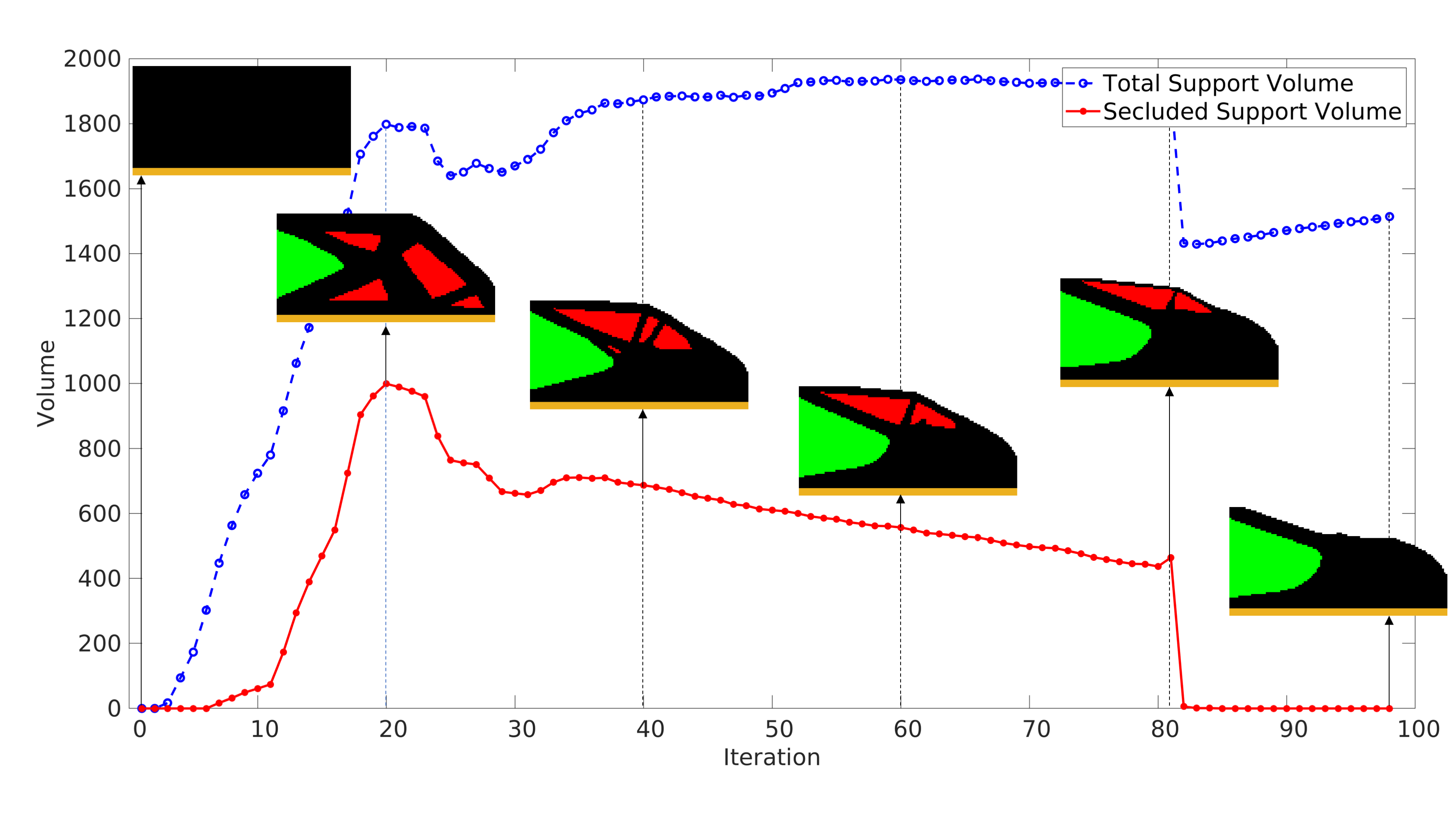}
	\caption{Volume of accessible and inaccessible support structure for cantilever beam TO of Fig. \ref{fig_benchmark2D}a.}
	\label{fig_supportConvergenceBeam2D}
\end{figure}

First, we consider a simple cantilever beam example in 2D and 3D. The loading 
conditions are shown in Fig. \ref{fig_beamLoading}. 
We use material properties of Stainless Steel with Young’s modulus of $E = 270 ~GPa$ and Poisson ratio of $\nu = 0.3$.
In 2D, we set the volume fraction to 0.5 and resolution at
 100$\times$50.  In 3D, the volume fraction is 0.3 and the resolution at 25$\times$51$\times$25.
In each example, we solve both accessibility-constrained and unconstrained TO and report the nonzero secluded support volume for the latter $V_{\Gamma_{unc}} > 0$ whose prevention comes at the cost of an increase in the resulting compliance values {\tiny }($\phi_{con} > \phi_{unc} $). An example of TO with no accessibility constraint on supports is shown in Fig. \ref{fig_mfgAnalysis_beam2D}a where the build direction is along +Z, and the milling tool can only approach the part from left ($\Theta = \{0\}$). Figures \ref{fig_mfgAnalysis_beam2D}b and \ref{fig_mfgAnalysis_beam2D}c illustrate the overhang points and support structures of the optimized design with $90 ^\circ$ overhang angle, meaning each point needs to be supported by some material directly underneath it. Figure \ref{fig_mfgAnalysis_beam2D}d shows the normalized IMF, which is used to identify accessible and secluded support structures depicted in Fig. \ref{fig_mfgAnalysis_beam2D}e. Observe that the platform is also considered as an obstacle. Figure \ref{fig_mfgAnalysis_beam2D}f shows the filtered IMF  $\bar{\mathcal{S}}$ defined in \eq{eq_accFilter}.
The 2D results for the cantilever beam example under different build and tool orientations are illustrated in Fig. \ref{fig_benchmark2D}. Performance and support accessibility results are summarized in Table \ref{tab_beam2Dresults}. As the number of tool orientations increase, the performance typically improves. However, for design of Fig. \ref{fig_benchmark2D}g the compliance is slightly higher than Fig. \ref{fig_benchmark2D}a and \ref{fig_benchmark2D}e, since in 2D the optimizer cannot exploit the higher degrees of freedom to generate a drastically different design and it converges to a slightly worse locally optimum design as the problem is not convex. Density and compliance convergence plots are shown in Figures \ref{fig_densityConvergenceBeam2D} and \ref{fig_complianceConvergenceBeam2D}, respectively. Observe that the support accessibility constraint is imposed after a few (e.g., 20) iterations and the graphs for with and without accessibility constraint coincide up to that point. As $w_{acc}$ gradually increases similar to continuation method widely used for density penalization, the constrained design deviates from the nominal unconstrained one to find a better performance (typically higher compliance or ${\varphi_{\con}}/{\varphi_{\unc}} \ge 1 $) with no secluded supports. Figure \ref{fig_supportConvergenceBeam2D} compares the values of total support volume and secluded support volume throughout the optimization. As expected, the support structure volume can drastically change from one design to another as overhang surfaces are created or removed.   

\begin{figure}[ht!]
	\centering
	\includegraphics[width=0.8\linewidth]{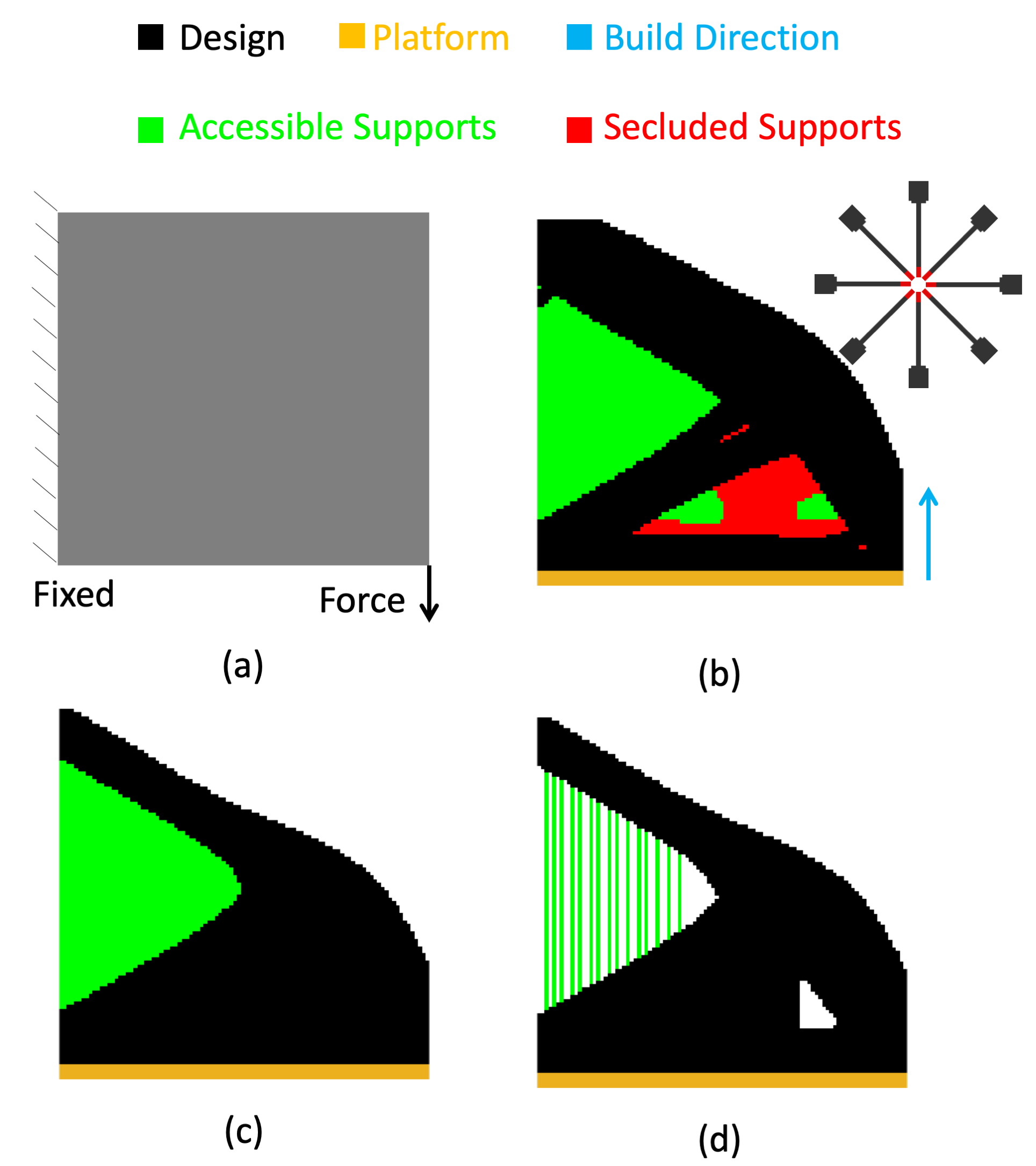}
	\caption{Difference of optimized designs for different overhang angles. (a) problem setup, (b) near-net optimized design without accessibility constraint, (c) near-net optimized design with accessibility constraint at $90^\circ$ overhang angle, and (d) near-net optimized design with accessibility constraint at $45^\circ$ overhang angle.}
	\label{fig_bechmarkOverhangAngle}
\end{figure}

\begin{figure} [ht!]
	\begin{subfigure}[t]{0.5\linewidth}
		\centering
		\includegraphics[width=\linewidth]{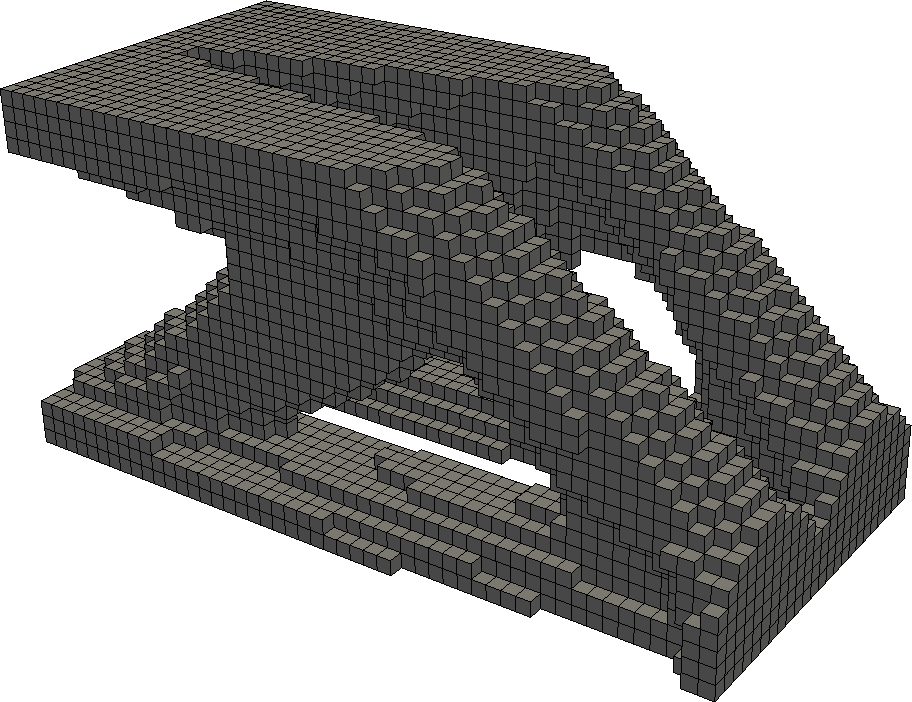}
		\caption{}
	\end{subfigure}%
	\begin{subfigure}[t]{0.5\linewidth}
		\centering
		\includegraphics[width=\linewidth]{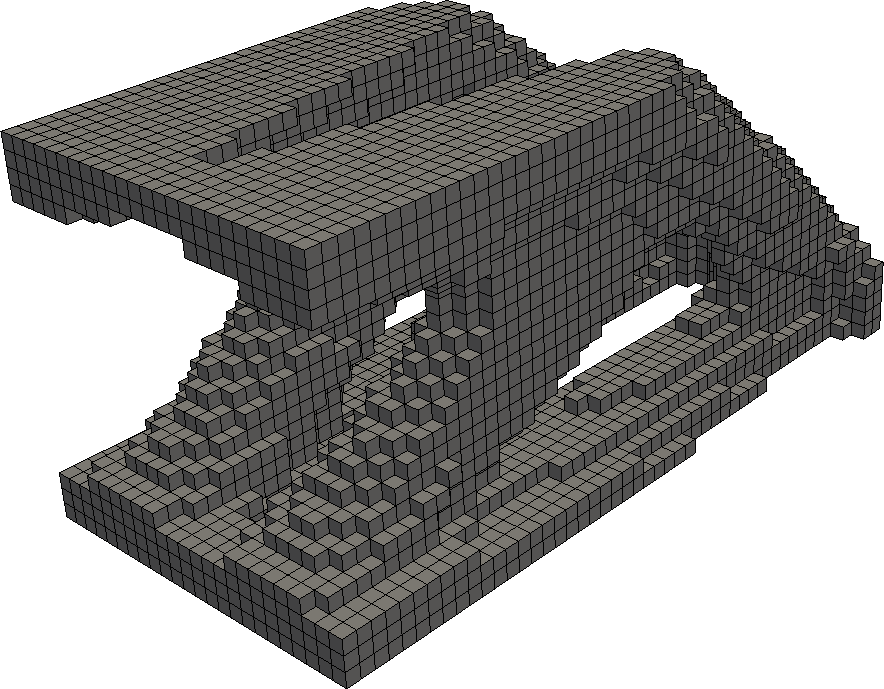}
		\caption{}
	\end{subfigure}%
	\caption{Voxel representation of the optimized cantilever beam with no support accessibility constraint at 0.3 volume fraction.} \label{fig_cantilever_voxel}
\end{figure}

The overhang angle in previous examples were conservatively chosen to be $90^\circ$. Let us consider the square cantilever beam example of Fig. \ref{fig_bechmarkOverhangAngle}a, where the build direction is along +Z and $[\Theta]= \{ R(\theta)_{} ~|~ \theta = 2\pi i/8 ~\text{for}~ i = 0, 1, \ldots, 7\}$. The unconstrained optimized design is shown in Fig. \ref{fig_bechmarkOverhangAngle}b. Observe that there are some isolated regions that are classified as accessible where there is no connected path for the cutting tool from outside of the part to that location. As was previously mentioned, the convolution-based IMF only checks for necessary condition of accessibility; that is whether the tool would collide with surrounding obstacles if cutter is placed at the point of interest. A sufficient condition would be to check for existence of a connected path for the tool to reach the point of interest. In the optimization, this usually is resolved in subsequent iterations as the surrounding secluded regions are penalized and become solid. In other words, as the boundary of the design evolves, these isolated accessible regions become secluded and correctly penalized until no such region exists, as illustrated in Fig. \ref{fig_bechmarkOverhangAngle}c with $\alpha = 90^\circ$ and Fig. \ref{fig_bechmarkOverhangAngle}d with $\alpha = 45^\circ$. Although we have not explicitly imposed any self-supporting constraint, the TO has automatically created a self-supporting enclosed void. 
Figure \ref{fig_cantilever_voxel} illustrates the voxel representation of the optimized cantilever beam example in 3D with no support accessibility constraint. 

\begin{figure}[ht!]
	\centering
	\includegraphics[width=\linewidth]{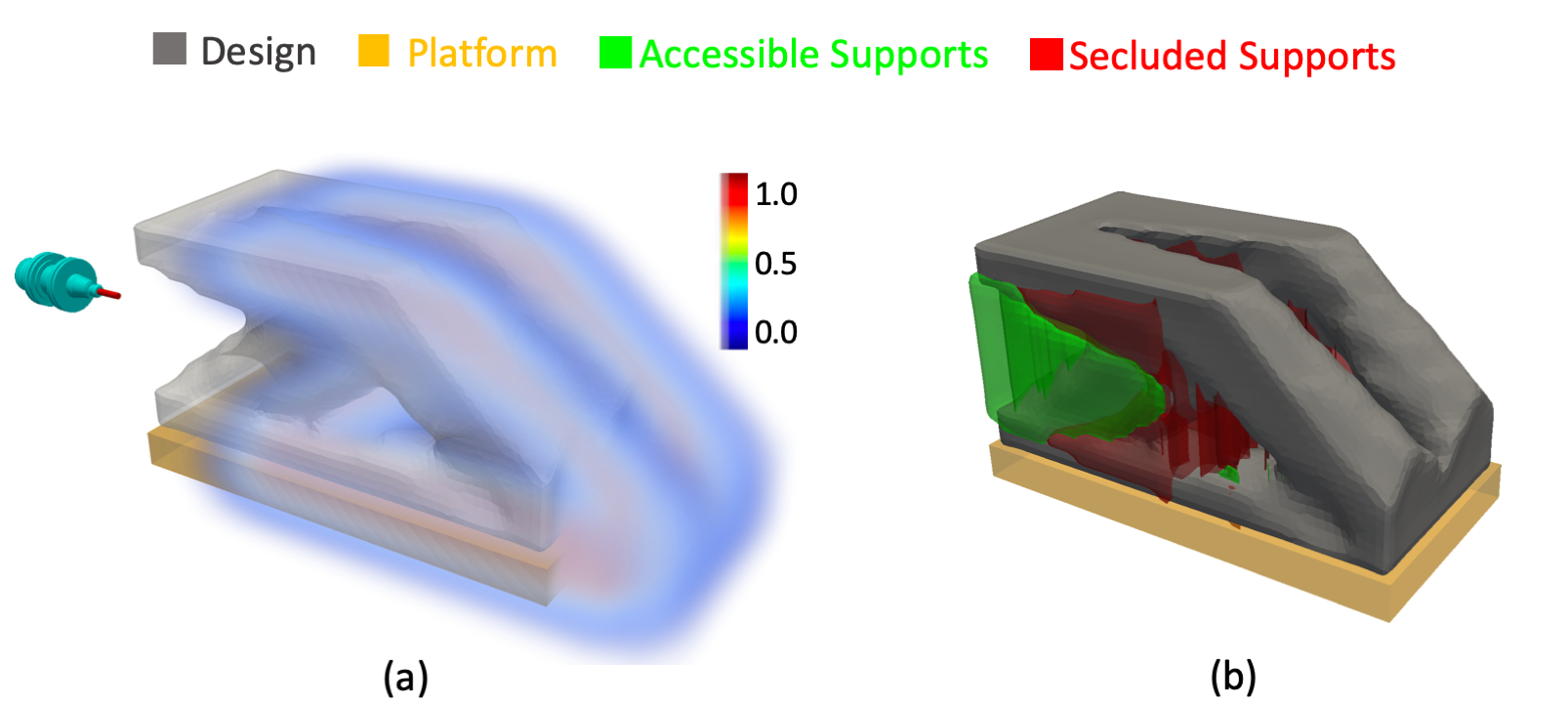}
	\caption{TO for cantilever beam in 3D at $0.3$ volume fraction \textit{without} accessibility constraint on support structures for $45^\circ$ overhang angle: (a) IMF by $0^\circ$ tool angle with $+Z$ build orientation, (b) accessible and secluded supports.}
	\label{fig_uncBeam3D}
\end{figure}

\begin{figure}[ht!]
	\centering
	\includegraphics[width=\linewidth]{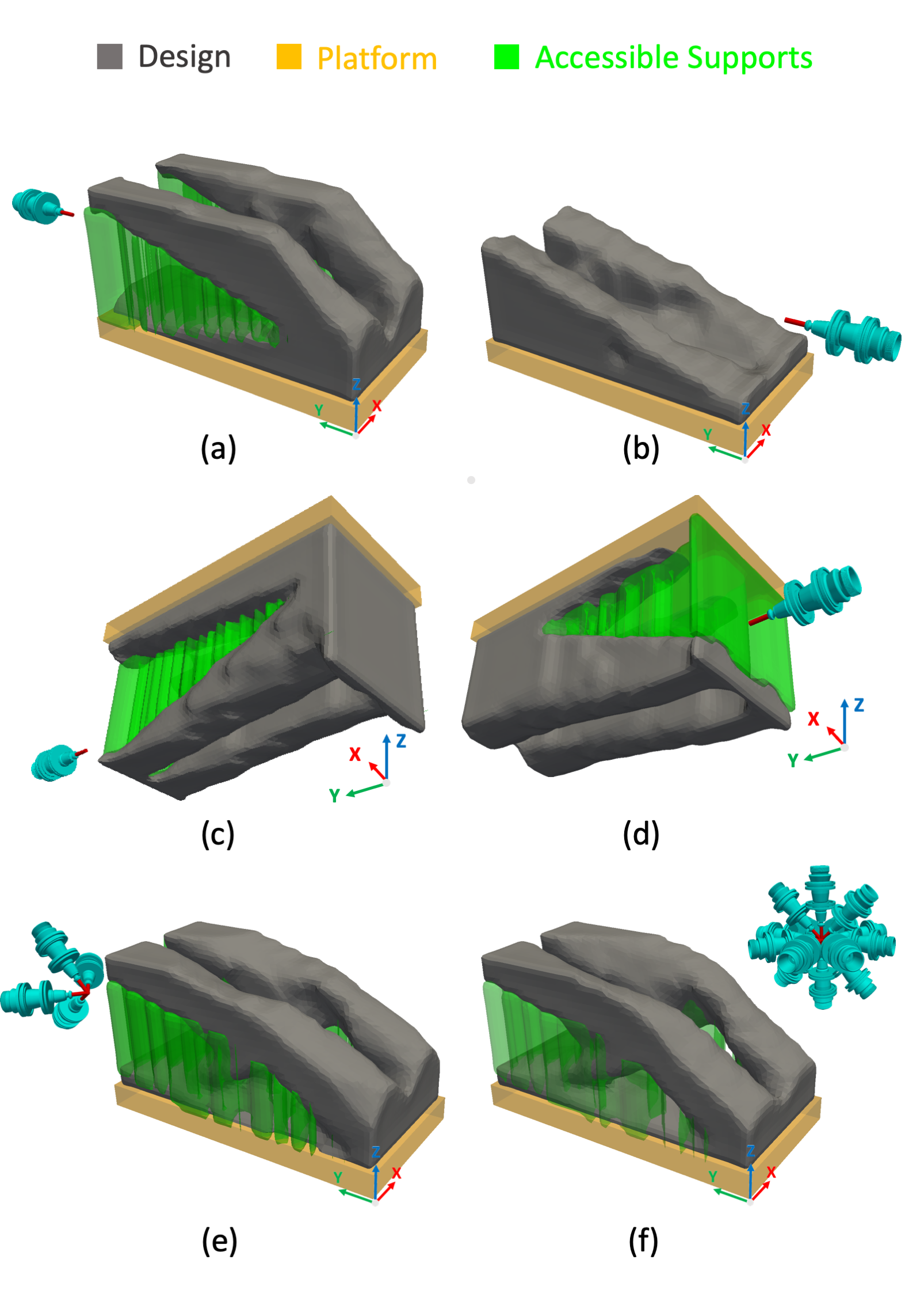}
	\caption{TO for cantilever beam in 3D \textit{with} accessibility constraint on support structures for $45^\circ$ overhang angle: (a) $R_X(0)$ tool angle with $+Z$ build orientation, (b) $R_X(\pi)$ tool angle with $+Z$ build orientation, (c) $R_X(0)$ tool angle with $-Z$ build orientation, (d) $R_X(\pi)$ tool angle with $-Z$ build orientation, (e) $\left\{  R_X(\pm \pi/4),R_Z(\pm \pi/4), \right\}$ tool angle with $+Z$ build orientation, (f) $\{ R_X(0), R_X(\pi), R_X(\pm \pi/2), R_X(\pm\pi/4),R_X(\pm3\pi/4), R_Z(\pm \pi/2),$\\ 
		$R_Z(\pm\pi/4),R_Z(\pm3\pi/4)\}$ tool angle with $+Z$ build orientation.}
	\label{fig_benchmark3D}
\end{figure}

Figure \ref{fig_uncBeam3D}a illustrates the optimized cantilever beam example in 3D with no support accessibility constraint and the corresponding IMF. The cutting tool is approaching the part from the left, build direction is along +Z, and the platform is also considered as an obstacle. Figure \ref{fig_uncBeam3D}b shows the optimized design with accessible and secluded support structures where  $\dfrac{V_{\Gamma_{unc}}}{V_{\mathsf{S}_{unc}}} = 0.43$. The 3D results for the cantilever beam example under different build and tool orientations are illustrated in Fig. \ref{fig_benchmark3D}. Performance and support accessibility results are summarized in Table \ref{tab_beam3Dresults}. Observe that in 3D, as the number of available tool orientations increase, the performance of the optimized design improves. In the case of Fig. \ref{fig_benchmark3D}f, there are 14 tool orientations and a very small amount of support in the unconstrained design is secluded ($\dfrac{V_{\Gamma_{unc}}}{V_{\mathsf{S}_{unc}}}$ is almost 0). Nonetheless, imposing the additional accessibility constraint slightly changes the design and compliance is slightly increased ($\dfrac{\varphi_{\con}}{\varphi_{\unc}} = 1.12$).

Table \ref{tab_time} provides a benchmark study on computational time for FEA, support generation, and IMF computation at different voxel resolutions. IMF is based on FFT algorithm that is implemented on GPU and is evaluated very efficiently (Fractions of a second). 
\begin{table} [ht!]
	\centering
	\caption{Summary of cantilever beam results in 2D.}
	\tabulinesep=1mm
	\begin{tabu}[t!]{cccc}
		\hline \hline 
		$b$ & $\Theta$ & $\dfrac{\varphi_{\con}}{\varphi_{\unc}}$ & $\dfrac{V_{\Gamma_{unc}}}{V_{\mathsf{S}_{unc}}}$ \\
		\hline
		$(0,+1)$ & $\{R(0)\}$ & 1.53 & 0.62 \\
		\hline
		$(0,+1)$ & $\{ R(\pi)\}$ & 2.66 & 1.00 \\
		\hline
		$(0,-1)$ & $\{R(0)\}$ & 1.93 & 0.72 \\
		\hline
		$(0,-1)$ & $\{ R(\pi)\}$  & 3.33 & 0.70 \\
		\hline
		$(0,+1)$ & $\{ R(0),R(\pi)\}$ & 1.37 & 0.62 \\
		\hline
		$(0,+1)$ & $\{ R(\pm \pi/4)\}$  & 1.89 & 0.74 \\
		\hline
		$(0,+1)$ &  {$\!\begin{aligned} %
				\{&R(0), R(\pi), R(\pm\pi/2),\\
				&R(\pm\pi/4),R(\pm 3\pi/4) \end{aligned}$}  & 1.62 & 0.62 \\
		\hline
	\end{tabu}
	\label{tab_beam2Dresults}
	\vspace{1.0em}
	\centering
	\caption{Summary of cantilever beam results in 3D.}
	\tabulinesep=1mm
	\begin{tabu}[t!]{cccc}
		\hline \hline 
		$b$ & $\Theta$ & $\dfrac{\varphi_{\con}}{\varphi_{\unc}}$ & $\dfrac{V_{\Gamma_{unc}}}{V_{\mathsf{S}_{unc}}}$ \\
	\hline
	$(0,0,+1)$ & $\{R_X(0)\}$ & 2.23 & 0.43 \\
	\hline
	$(0,0,+1)$ & $\{R_X(\pi)\}$ & 5.13 & 0.91 \\
	\hline
	$(0,0,-1)$ & $\{R_X(0)\}$ & 3.42 & 0.64 \\
	\hline
	$(0,0,-1)$ & $\{R_X(\pi)\}$  &  10.00 & 0.79 \\ 
	\hline
	$(0,0,+1)$ & $\{ R_X(\pm \pi/4),R_Z(\pm \pi/4)\}$  &  1.18 & 0.029 \\ \hline
	$(0,0,+1)$ &   {$\!\begin{aligned} %
	\{&R_{X}(0), R_{X}(\pi),\\
	&, R_{X}(\pm\pi/2),R_{X}(\pm\pi/4), \\
	&R_{X}(\pm 3\pi/4), R_{Z}(\pm\pi/2),\\
	&R_{Z}(\pm\pi/4),R_{Z}(\pm 3\pi/4)\}  \end{aligned}$} & 1.12 &  0.00\\
		\hline
	\end{tabu}
	\label{tab_beam3Dresults}
\end{table}

\begin{table} [!t]
	\centering
	\caption{Computational time of FEA, support generation, and IMF computation.}
	\tabulinesep=0.5mm
	\begin{tabu}[t!]{cccc}
		\hline \hline 
		Resolution & FEA (s) & Supp. Gen. (s) & IMF (s) \\  
		\hline 
		 25$\times$51$\times$25 & 1.74 & 0.0006 & 0.009 \\
		 50$\times$100$\times$50 & 12.13 & 0.002 & 0.05 \\
		 75$\times$150$\times$75 & 39.9 & 0.007 & 0.1 \\
		\hline
	\end{tabu}
	\label{tab_time}
\end{table}

\subsection{GE Bracket}
\begin{figure}[ht!]
	\centering
	\includegraphics[width=\linewidth]{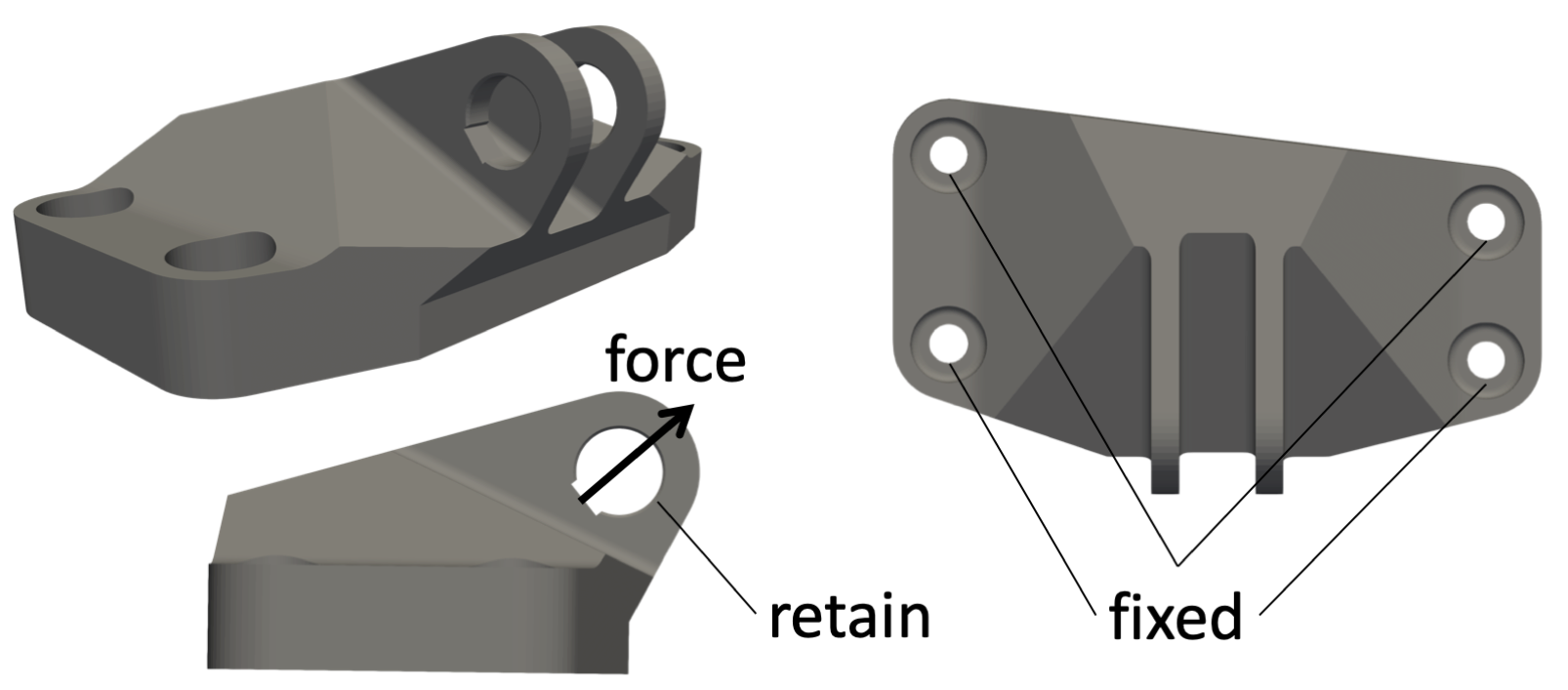}
	\caption{GE bracket geometry and boundary conditions.}
	\label{fig_GELoading}
\end{figure}

\begin{figure} [ht!]
	\begin{subfigure}[t]{\linewidth}
		\centering
		\includegraphics[width=\linewidth]{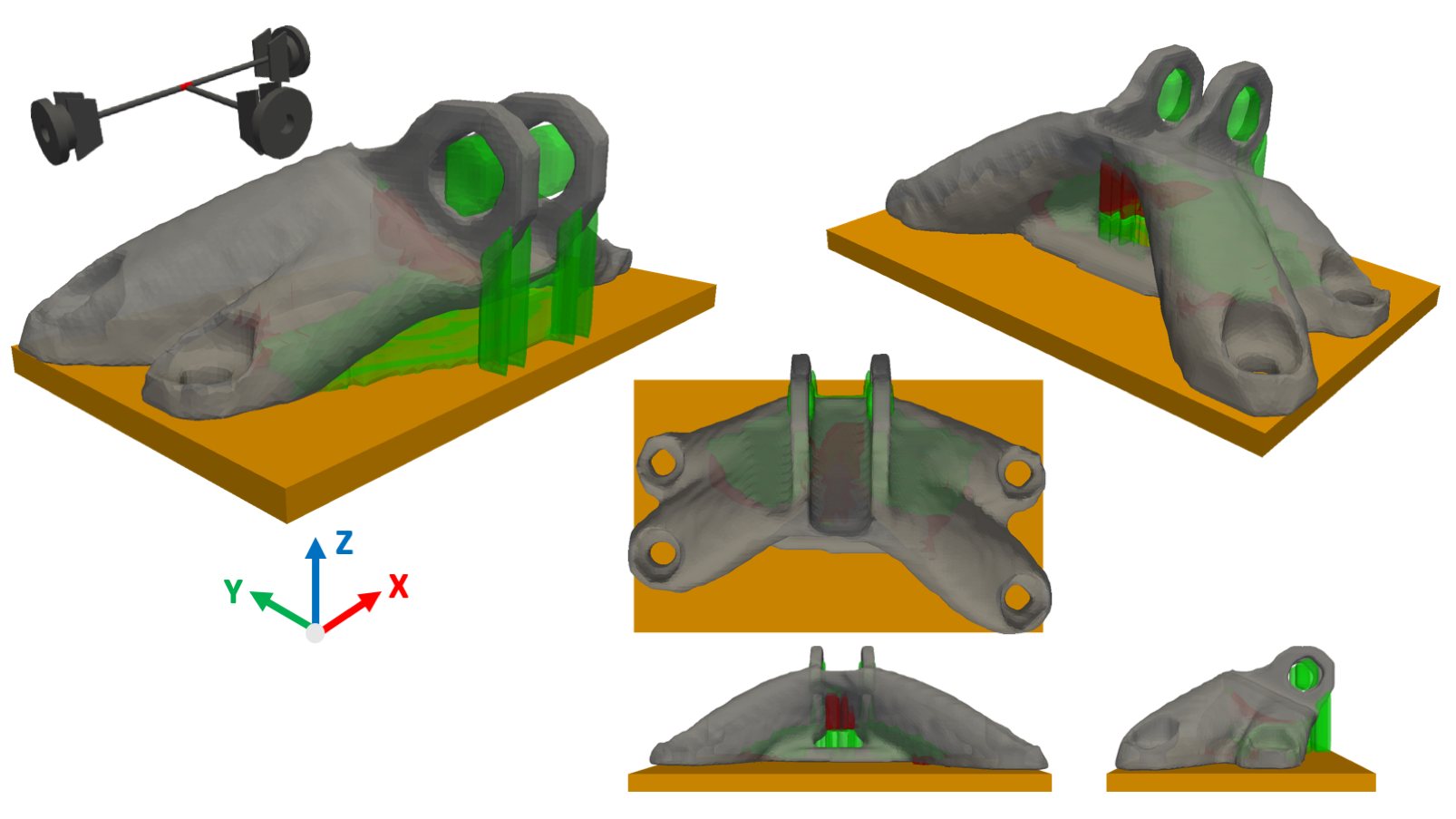}
		\caption{Without accessibility constraint.}
	\end{subfigure}%
	\\
	\begin{subfigure}[t]{\linewidth}
		\centering
		\includegraphics[width=\linewidth]{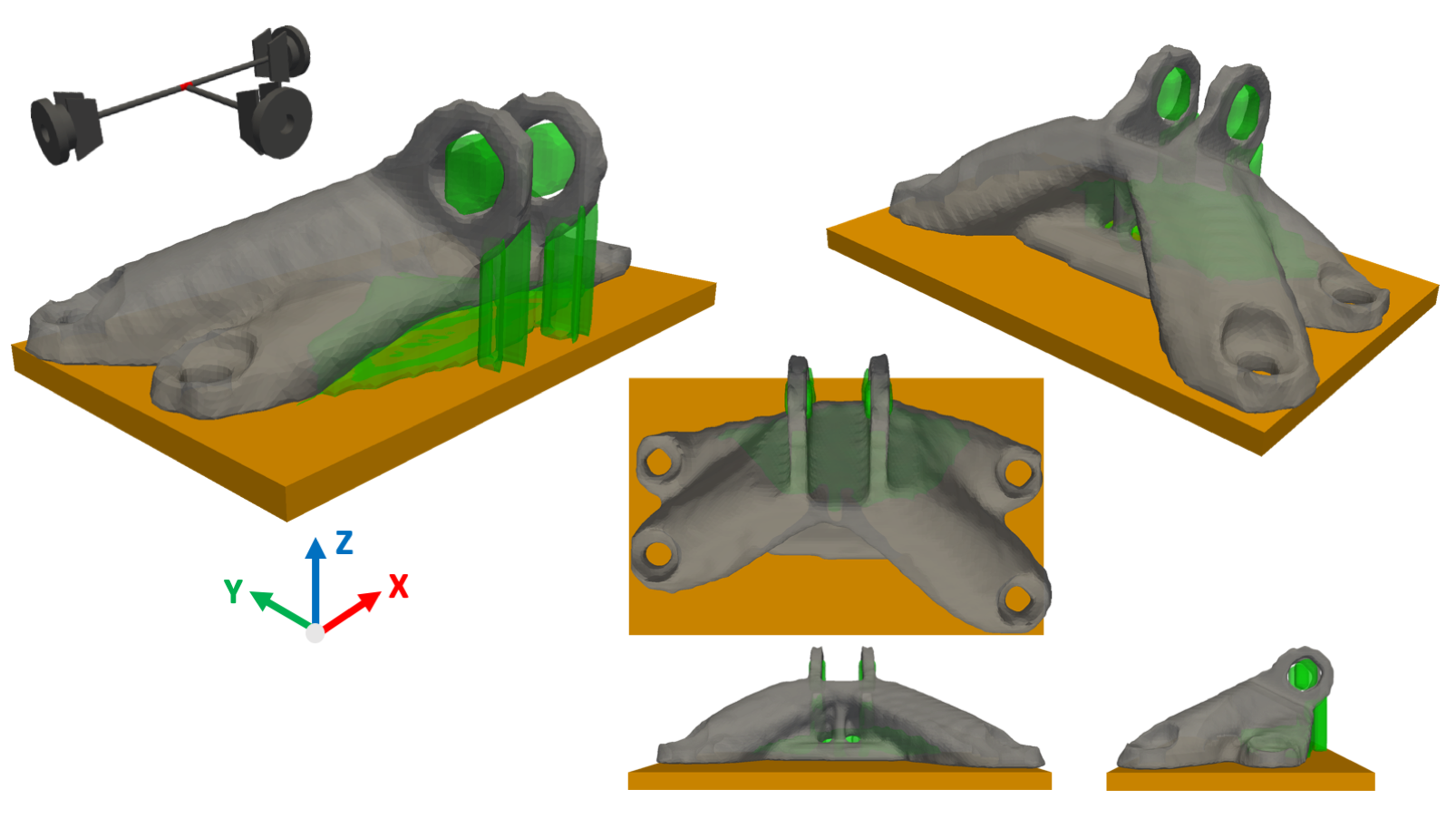}
		\caption{With accessibility constraint.}
	\end{subfigure}%
	\caption{Optimized GE bracket at 0.3 volume fraction and $+Z$ build direction.} \label{fig_GEbracket_2}
\end{figure}

\begin{figure} [ht!]
	\begin{subfigure}[t]{\linewidth}
		\centering
		\includegraphics[width=\linewidth]{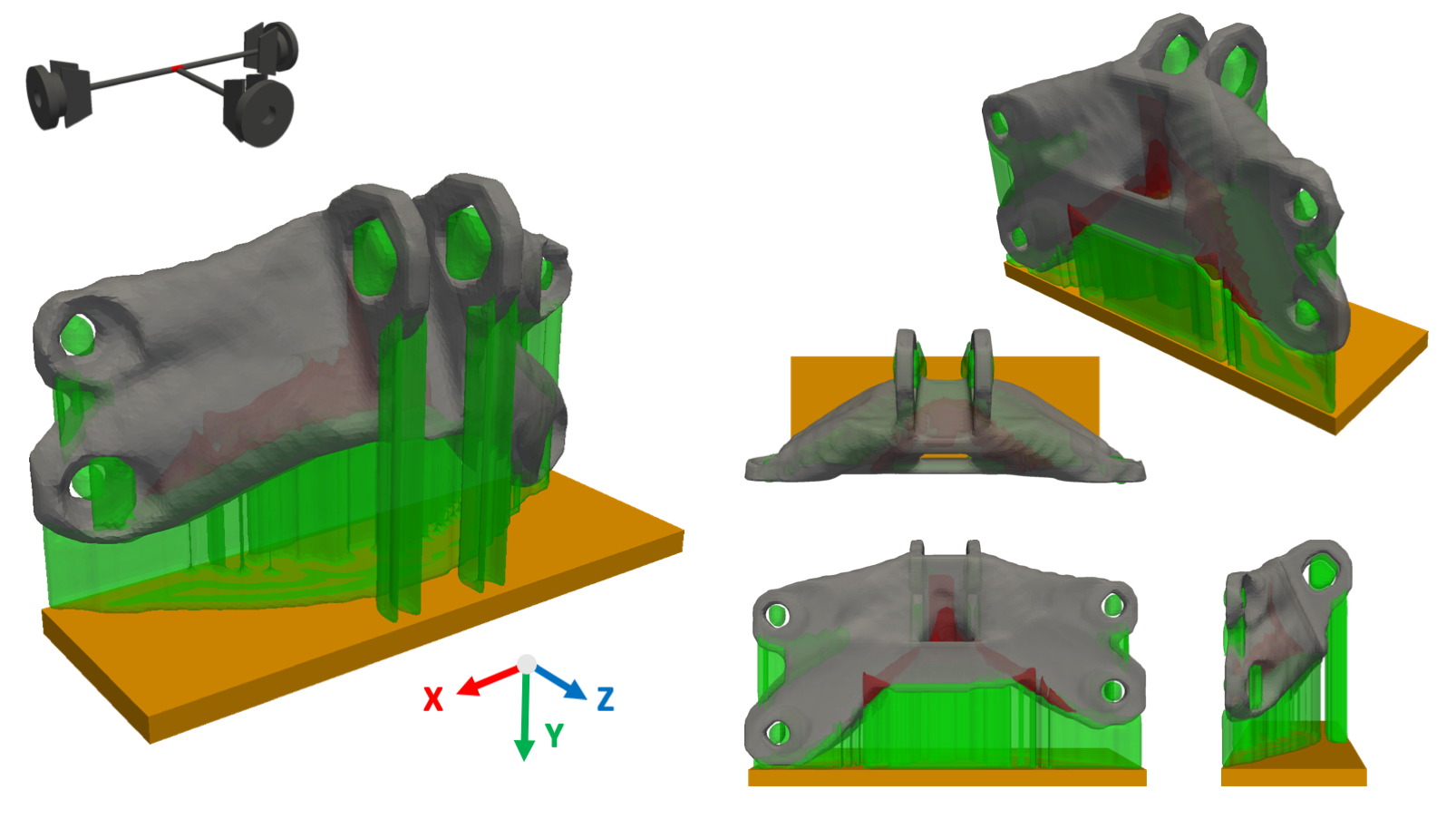}
		\caption{Without accessibility constraint.}
	\end{subfigure}%
	\\
	\begin{subfigure}[t]{\linewidth}
		\centering
		\includegraphics[width=\linewidth]{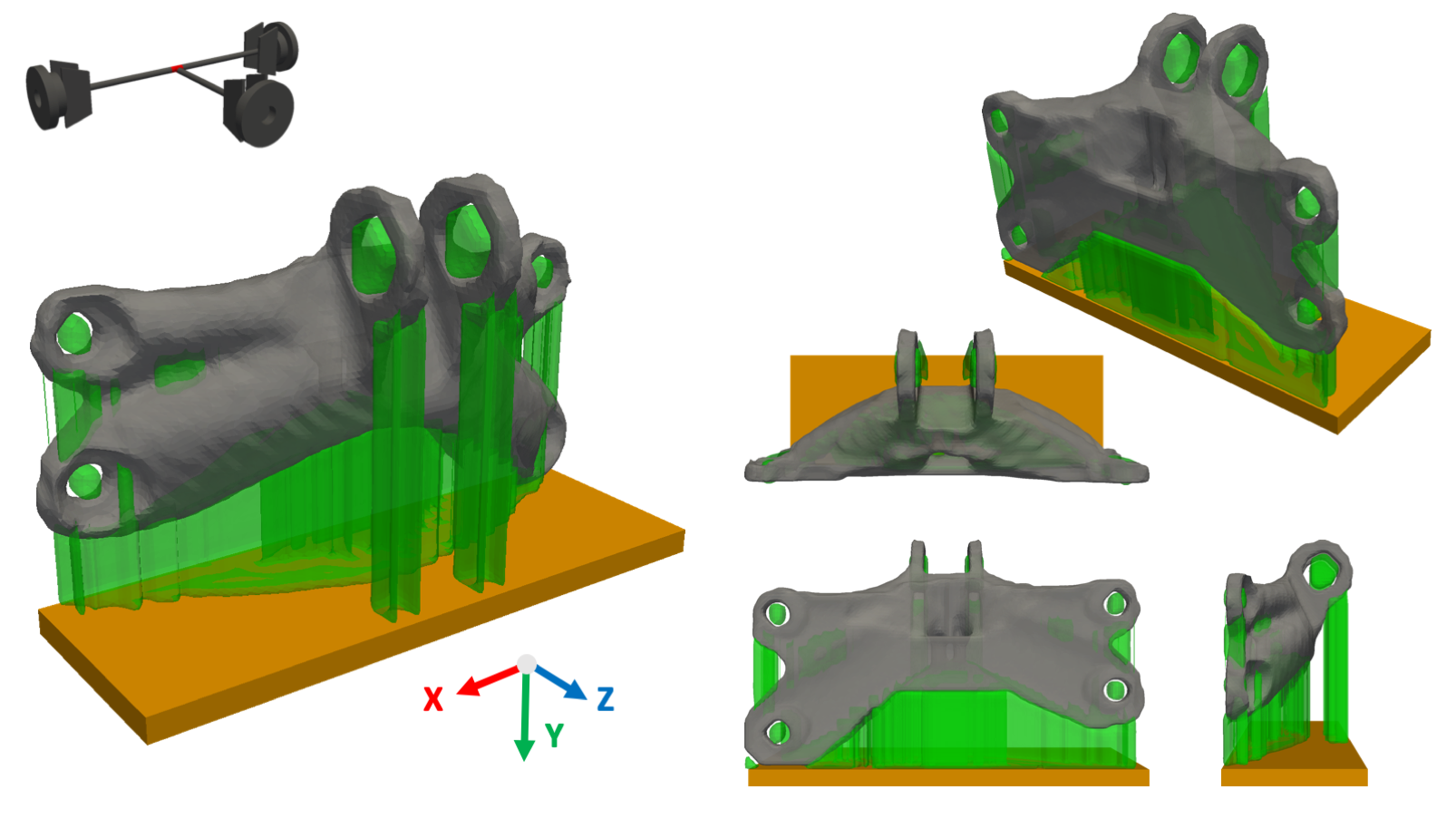}
		\caption{With accessibility constraint.}
	\end{subfigure}%
	\caption{Optimized GE bracket at 0.3 volume fraction and $-Y$ build direction.} \label{fig_GEbracket_1}
\end{figure}

\begin{figure} [ht!]
	\begin{subfigure}[t]{\linewidth}
		\centering
		\includegraphics[width=\linewidth]{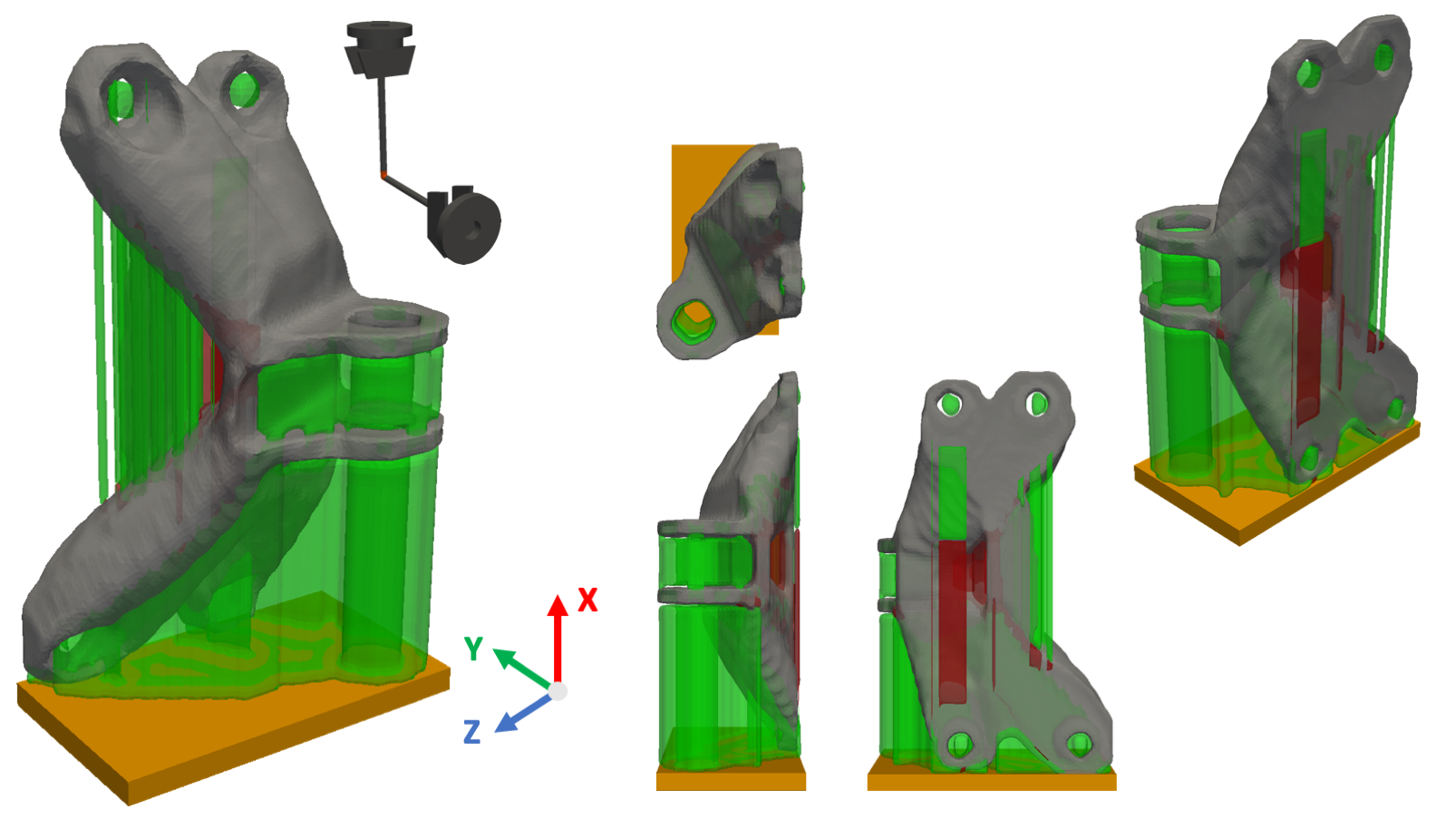}
		\caption{Without accessibility constraint.}
	\end{subfigure}%
	\\
	\begin{subfigure}[t]{\linewidth}
		\centering
		\includegraphics[width=\linewidth]{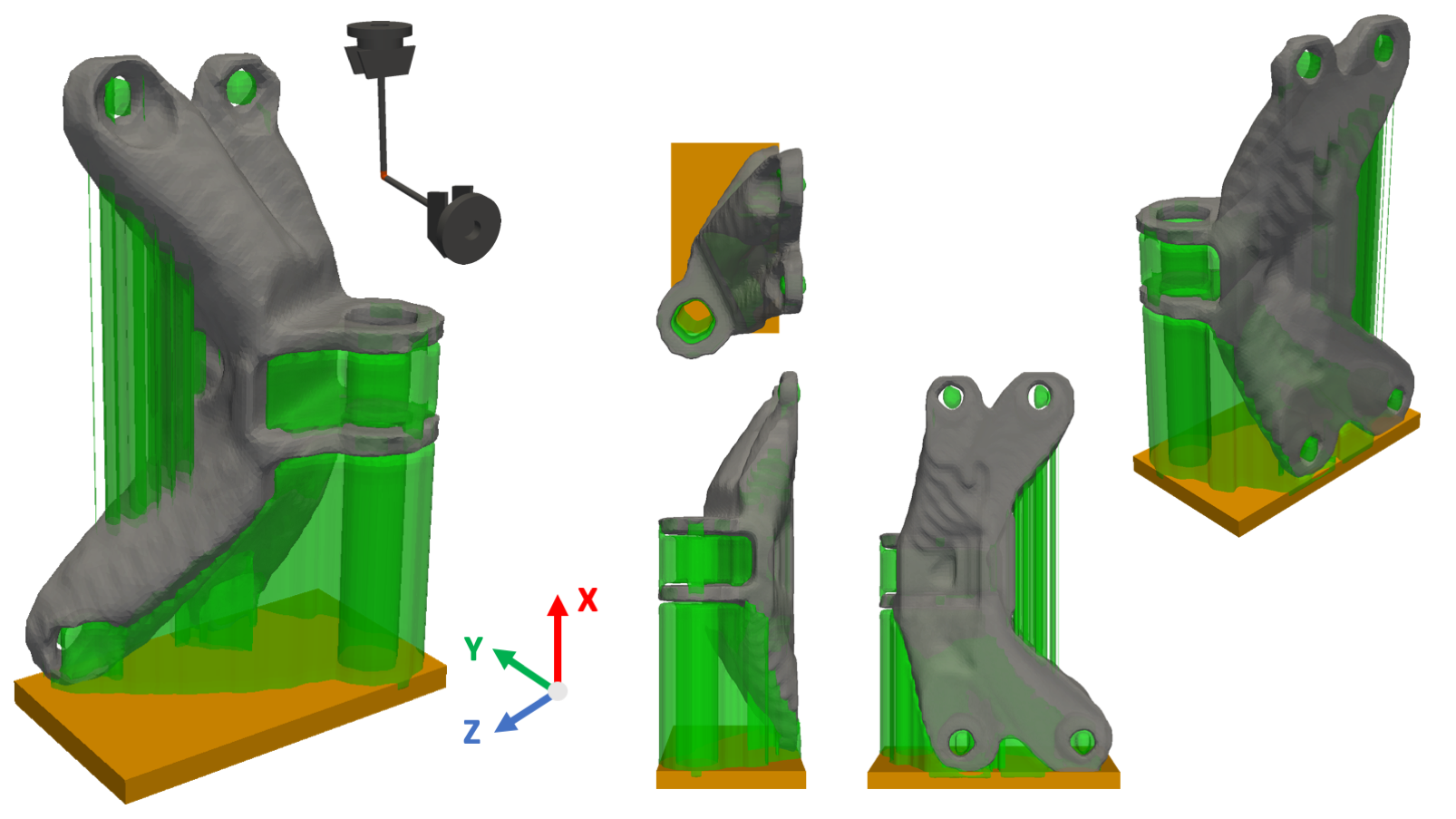}
		\caption{With accessibility constraint.}
	\end{subfigure}%
	\caption{Optimized GE bracket at 0.3 volume fraction and $+X$ build direction.} \label{fig_GEbracket_3}
\end{figure}

Next, let us consider the GE-bracket problem of Fig. \ref{fig_GELoading}. The material is similar to the previous example. The target volume fraction is 0.3 and the TO resolution is about 80,000 hexahedral finite elements. In all examples, we consider the build platform as an additional obstacle in IMF calculation.\\
Figure \ref{fig_GEbracket_2}a depicts the optimized design without support accessibility constraint built in +Z direction with about 9\% of supports being secluded. The optimized design with accessibility constraint is shown in Fig. \ref{fig_GEbracket_2}b with no secluded supports, while compliance increases about 11\% and the overall support ratio $\dfrac{V_{\mathsf{S}_{con}}}{V_{\mathsf{S}_{unc}}} = 0.65$.
Figure \ref{fig_GEbracket_1} illustrates the optimized bracket without and with accessibility constraint with build direction along -Y and 3 tool orientations. The design with no secluded support has 10\% higher compliance value than the unconstrained case and $\dfrac{V_{\mathsf{S}_{con}}}{V_{\mathsf{S}_{unc}}} = 0.74$. Figure \ref{fig_GEbracket_3} illustrated the optimized design build along +X direction with two tool orientations. Imposing accessibility constraint results in 10\% increase in compliance while the overall support volume reduces by about 7.5\%. Results for GE bracket examples are summarized in Table \ref{tab_GEresults}.
     
\begin{table} [ht!]
	\centering
	\caption{Summary of GE bracket results.}
	\tabulinesep=1mm
	\begin{tabu}[t!]{cccc}
		\hline \hline 
		$b$ & Tool Directions & $\dfrac{\varphi_{\con}}{\varphi_{\unc}}$ & $\dfrac{V_{\Gamma_{unc}}}{V_{\mathsf{S}_{unc}}}$ \\
		\hline
		$(0,0,+1)$ &  $\begin{dcases}
		(0,+1,0)\\
		(+1,0,0)\\
		(-1,0,0)
		\end{dcases}$ & 1.11 & 0.09 \\
		$(0,-1,0)$ & $\begin{dcases}
		(0,0,-1)\\
		(+1,0,0)\\
		(-1,0,0)
		\end{dcases}$ & 1.10 & 0.06\\
		$(+1,0,0)$ & $\begin{dcases}
		(0,+1,0)\\
		(-1,0,0)
		\end{dcases}$ & 1.10 & 0.05 \\
		\hline
	\end{tabu}
	\label{tab_GEresults}
\end{table}

\subsection{Quad-Copter}

\begin{figure}[ht!]
	\centering
	\includegraphics[width=0.8\linewidth]{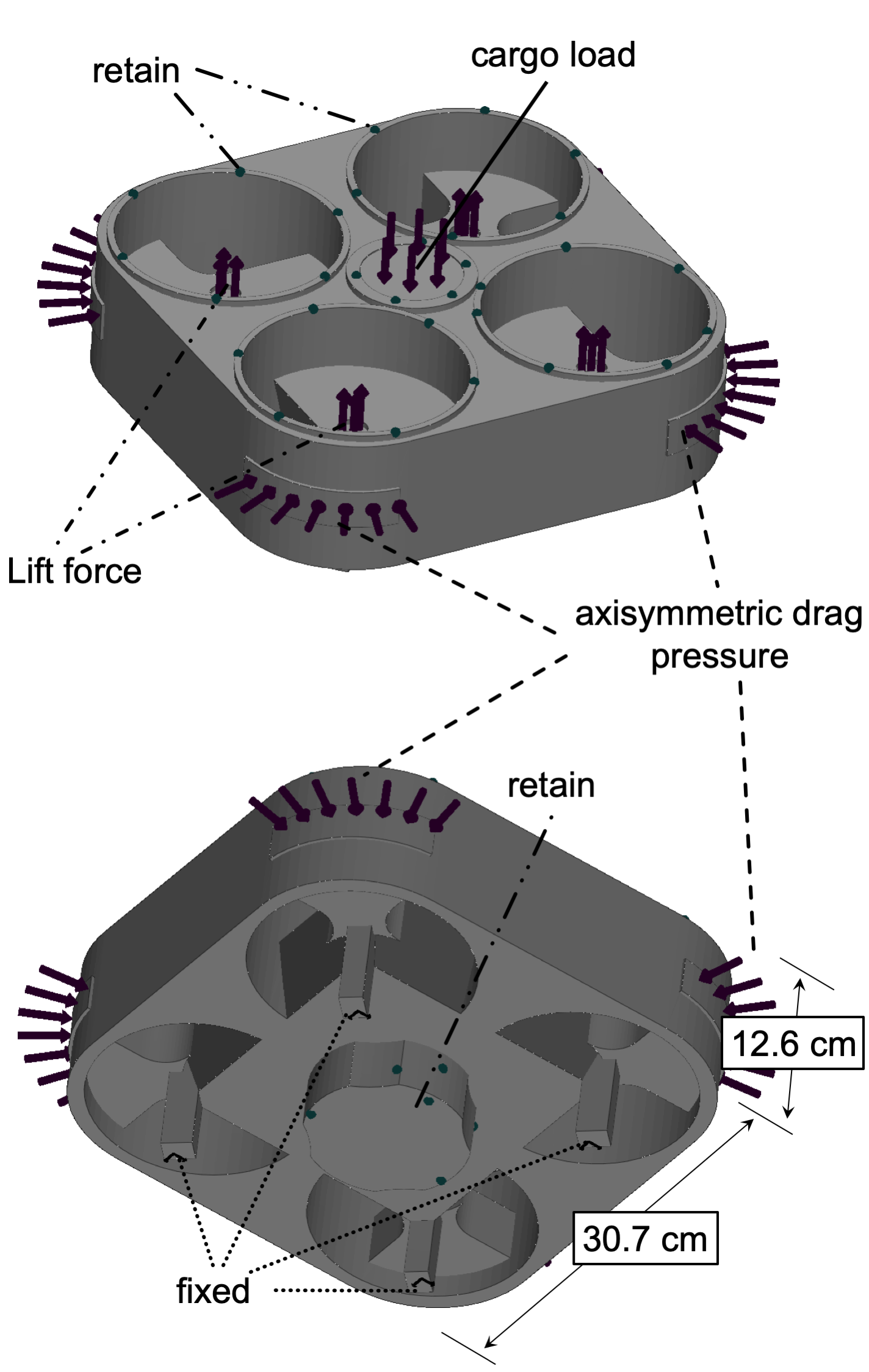}
	\caption{Quad-copter geometry and boundary conditions.}
	\label{fig_droneLoading}
\end{figure}

Figure \ref{fig_droneLoading} illustrates the geometry and simplified loading condition on a quad-copter considering lift and drag forces while retaining some battery-housing case and upper frame of the copter. The underlying material is assumed to be Aluminum 7075 with Young's modulus $E = 70 GPa$ and $\nu = 0.33$. The target volume fraction is 0.35 and the design domain is discretized into about 200,000 hexahedral elements.

\begin{figure}[ht!]
	\begin{subfigure}[t]{\linewidth}
		\centering
		\includegraphics[width=\linewidth]{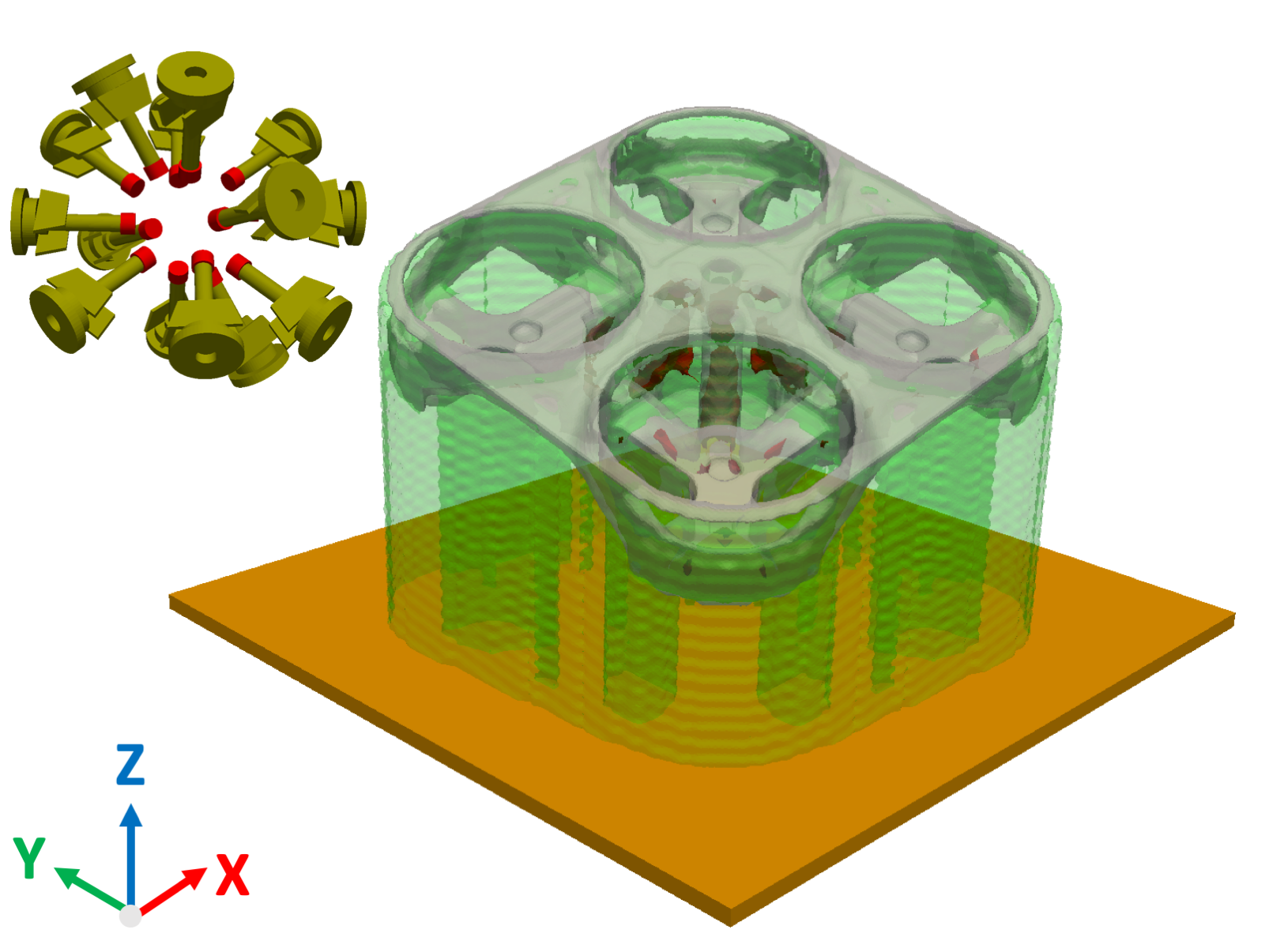}
		\caption{Without accessibility constraint.}
	\end{subfigure}%
	\\
	\begin{subfigure}[t]{\linewidth}
		\centering
		\includegraphics[width=\linewidth]{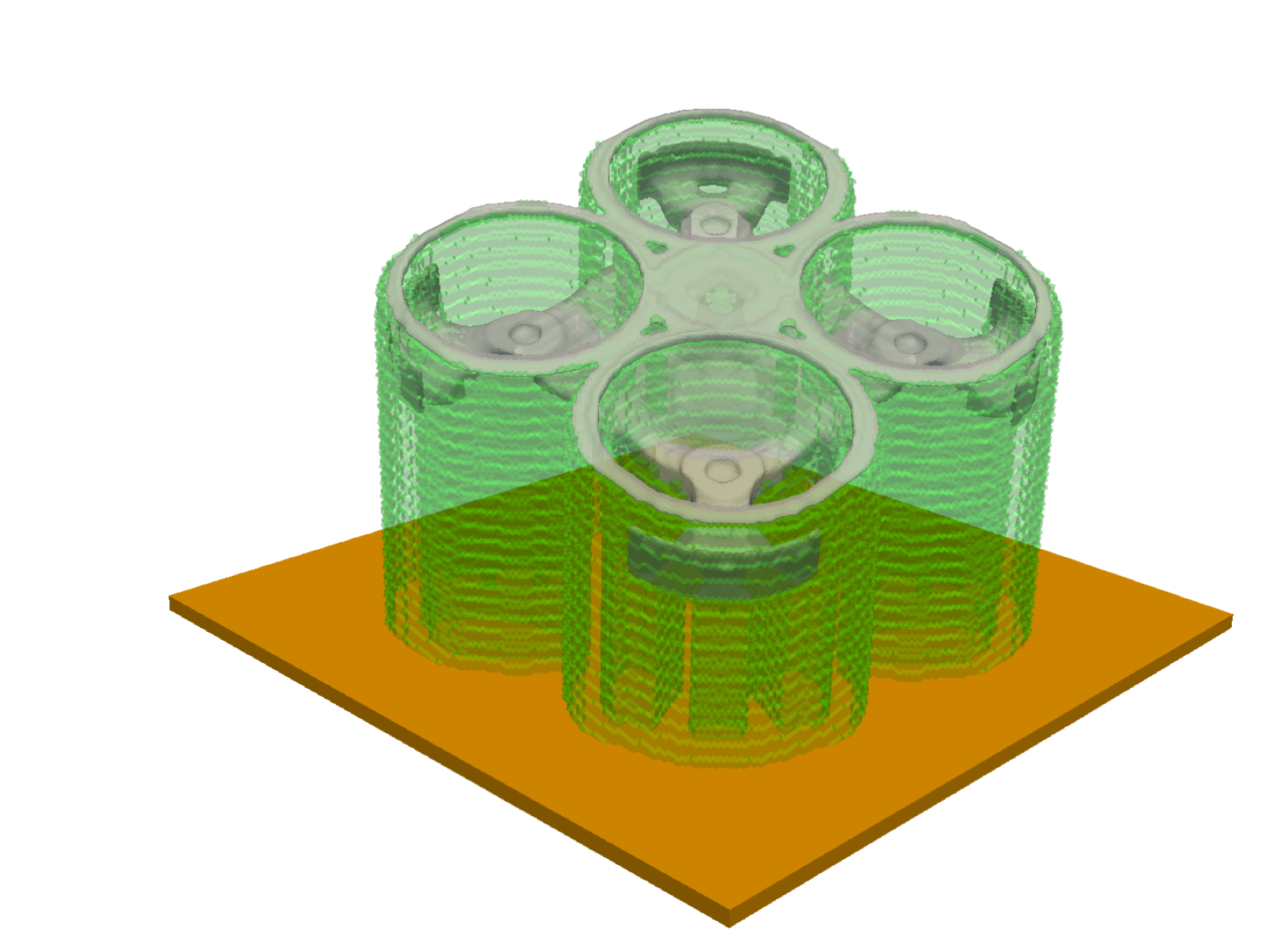}
		\caption{With accessibility constraint.}
	\end{subfigure}%
	\caption{Optimized quad-copter at 0.35 volume fraction and multi-axis milling tool. (a) \textit{without} accessibility constraint, and (b) \textit{with} accessibility constraint.} \label{fig_quadCopterComparison}
\end{figure}

\begin{figure}[ht!]
	\begin{subfigure}[t]{\linewidth}
		\centering
		\includegraphics[width=\linewidth]{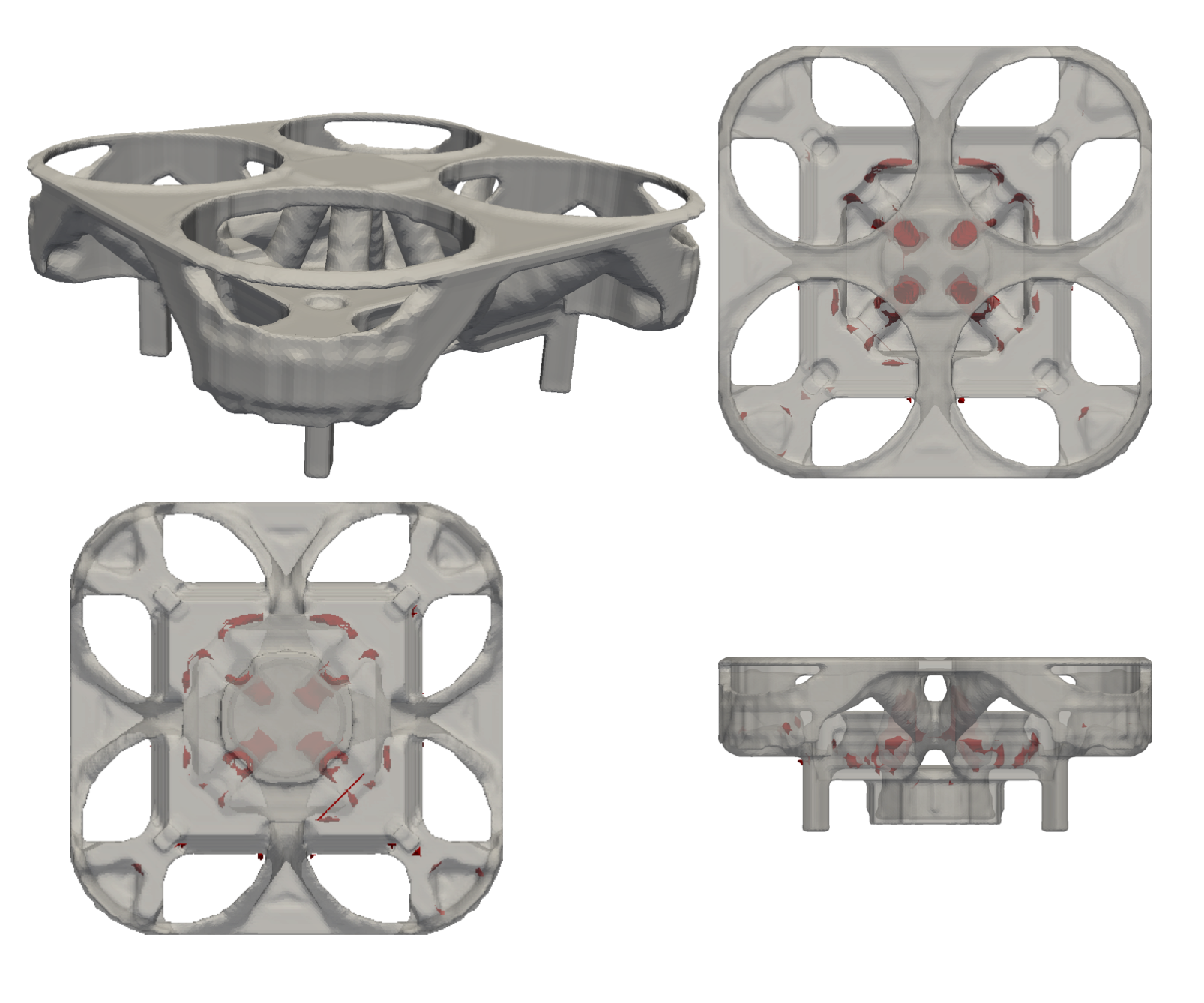}
		\caption{Without accessibility constraint.}
	\end{subfigure}%
	\\
	\begin{subfigure}[t]{\linewidth}
		\centering
		\includegraphics[width=\linewidth]{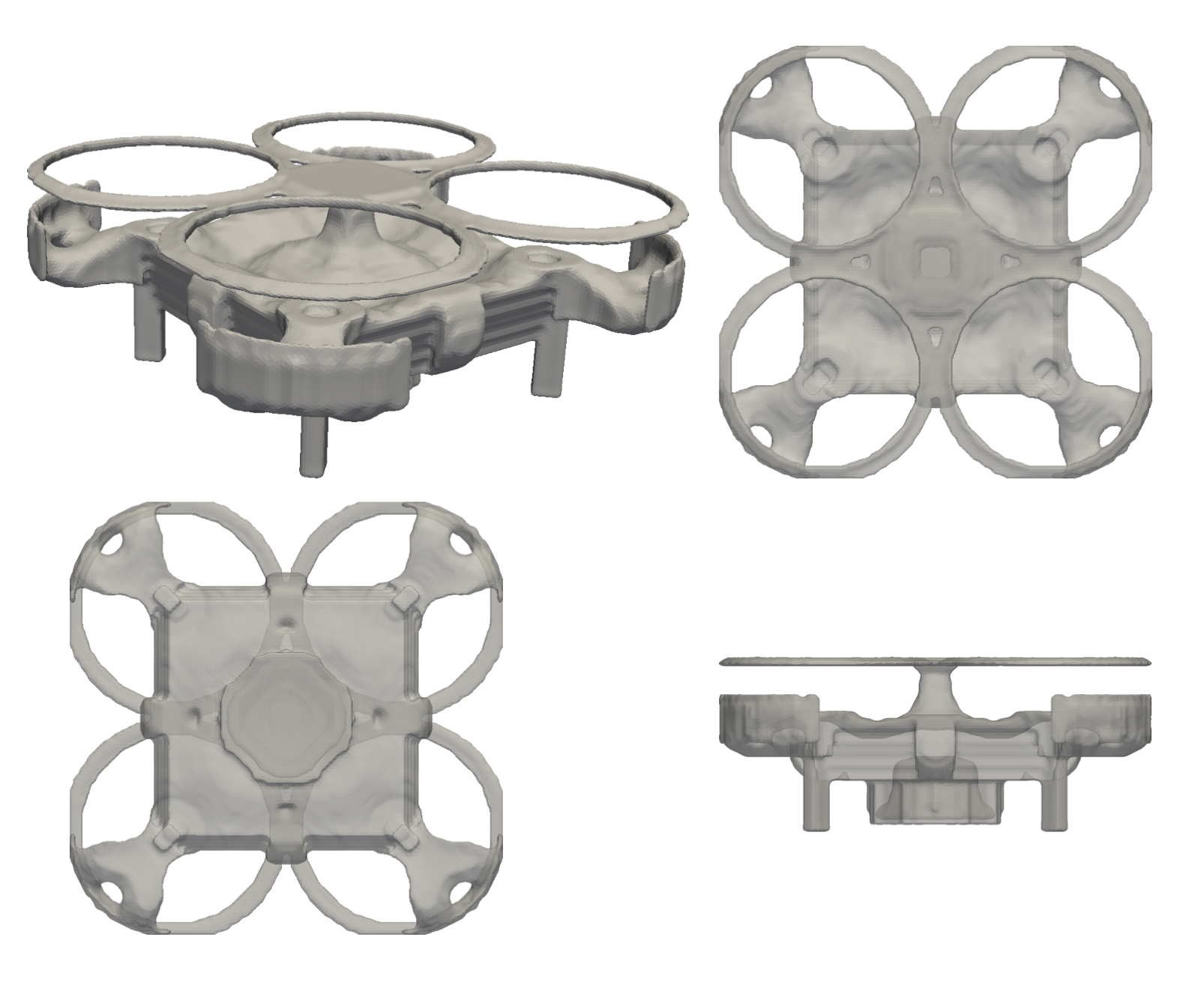}
		\caption{With accessibility constraint.}
	\end{subfigure}%
	\caption{Optimized quad-copter bracket at 0.35 volume fraction.} \label{fig_quadCopter}
\end{figure}

The optimized result without support accessibility constraint is shown in Fig. \ref{fig_quadCopterComparison}, where the build direction $\textbf{b} = \left\{0.3,0.3,1.0\right\}$ and the set of 14 tool orientations: 
\begin{align} %
	\Theta = \{&R_{X}(0), R_{X}(\pi),\nonumber\\
	&, R_{X}(\pm\pi/2),R_{X}(\pm\pi/4),\nonumber \\
	&R_{X}(\pm 3\pi/4), R_{Z}(\pm\pi/2),\nonumber\\
	&R_{Z}(\pm\pi/4),R_{Z}(\pm 3\pi/4)\} \nonumber  
\end{align}
Despite a large volume of seemingly unnecessary supports, this setup ensures that the initial design domain is manufacturable, i.e., all supports are accessible.  Figure \ref{fig_quadCopterComparison}a depicts the optimized design, accessible supports, and secluded supports in the absence of support constraint where the volume of secluded supports is about 35 $cm^3$. Figure \ref{fig_quadCopterComparison}b illustrates the optimized design with accessibility constraint, where there is no secluded support region. The compliance values between the two scenarios are quite comparable, $\dfrac{\varphi_{\con}}{\varphi_{\unc}} = 1.02$. Figure \ref{fig_quadCopter} illustrated different views of the final designs with secluded support region.       

\subsection{Upright}

\begin{figure}[ht!]
	\centering
	\includegraphics[width=\linewidth]{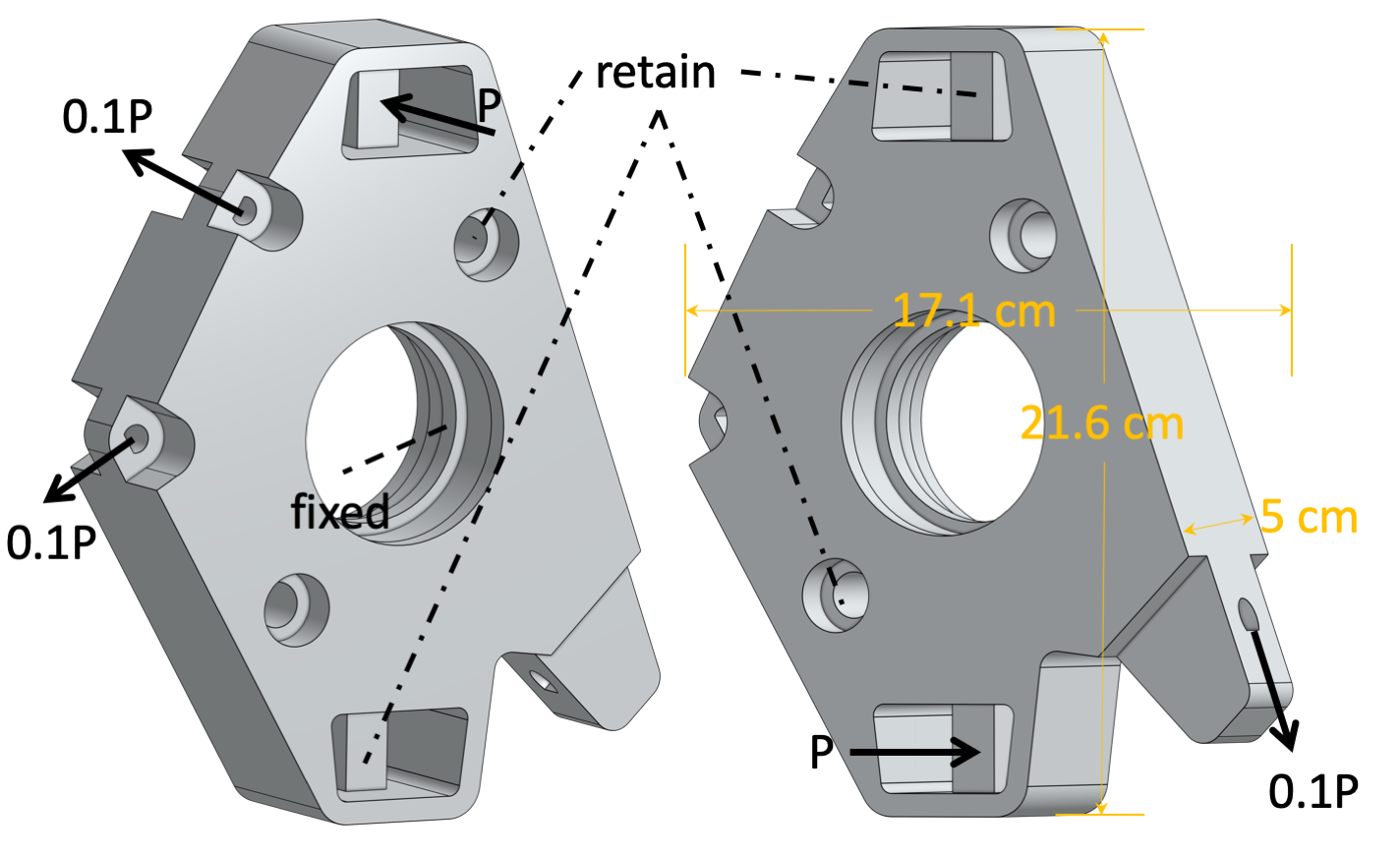}
	\caption{Upright geometry and boundary conditions.}
	\label{fig_uprightLoading}
\end{figure}

Finally, let us consider the upright example of Fig. \ref{fig_uprightLoading} with relative pressure values applied on different surfaces. The material properties are those of Ti6Al4V, with $E = 113.8$ GPa and $\nu = 0.34$. The design domain is discretized into about 100,000 hexahedral finite elements. 

\begin{figure} [ht!]
	\begin{subfigure}[t]{\linewidth}
		\centering
		\includegraphics[width=\linewidth]{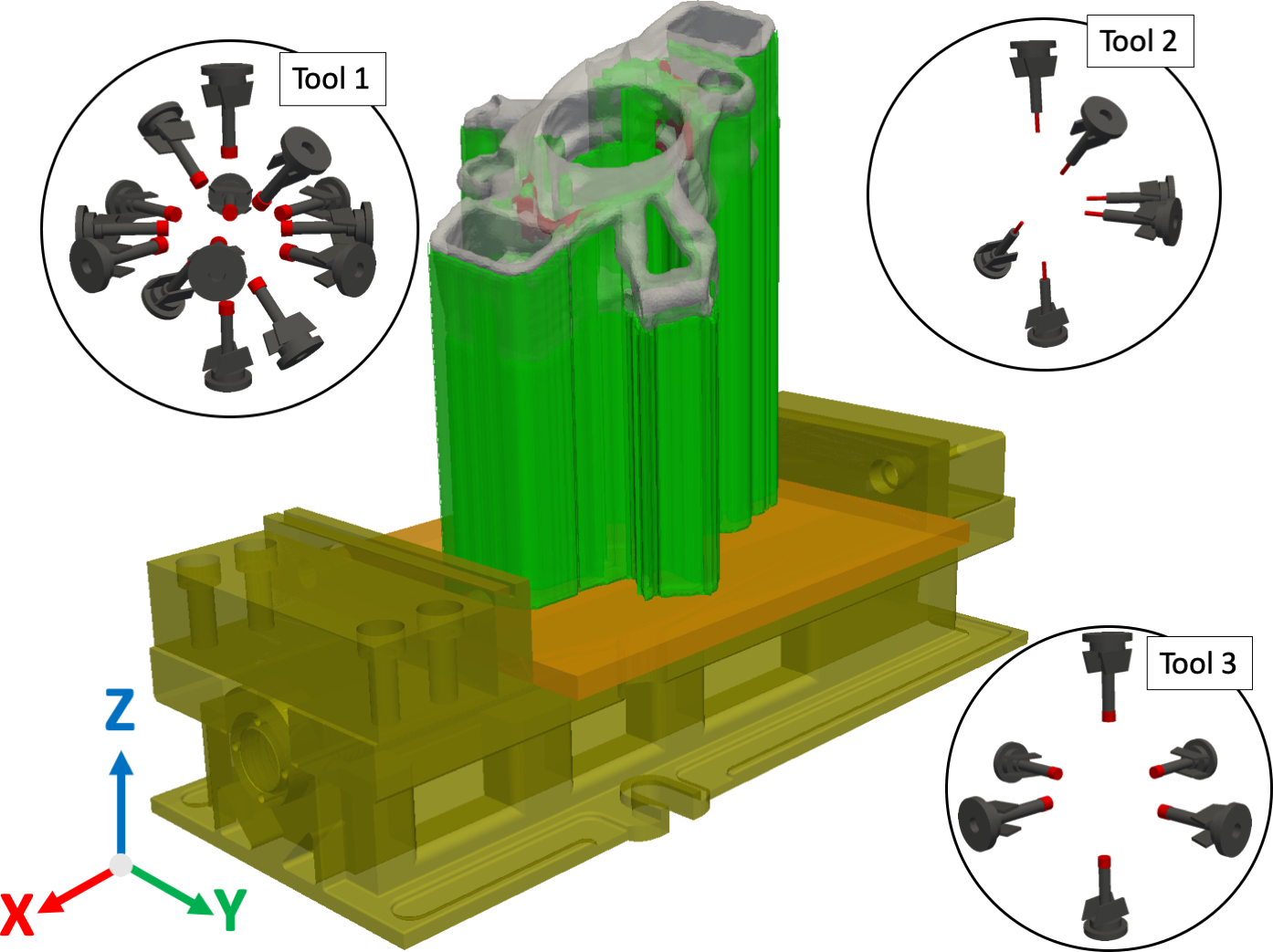}
		\caption{Without accessibility constraint.}
	\end{subfigure}%
	\\
	\begin{subfigure}[t]{\linewidth}
		\centering
		\includegraphics[width=0.8\linewidth]{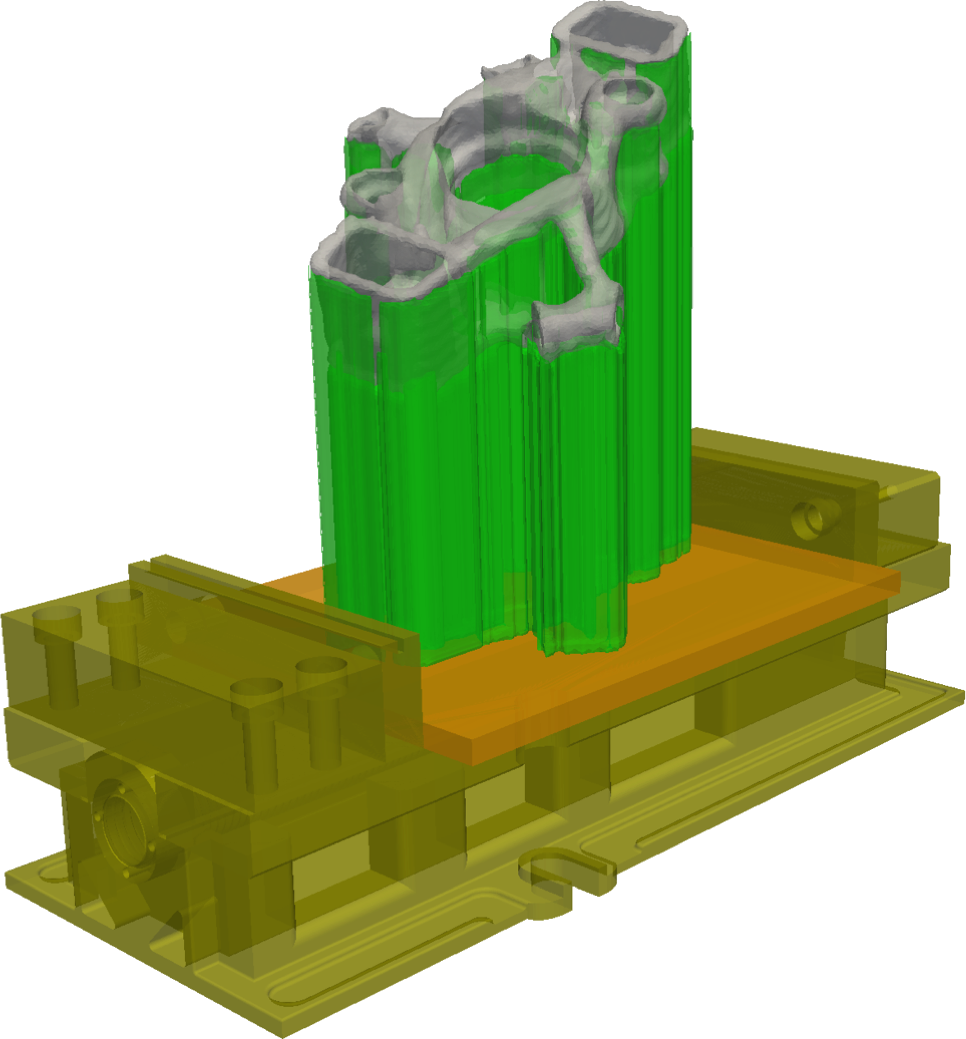}
		\caption{With accessibility constraint.}
	\end{subfigure}%
	\caption{Optimized upright bracket at 0.35 volume fraction, with  build direction of (0.25, 0.25,0.85), VISE fixture holding the build platform, and three milling tools and 26 tool orientations (tool 1: 14, tool 2: 6, and tool 3: 6 orientations).} \label{fig_upright_supp}
\end{figure}

\begin{figure} [ht!]
	\begin{subfigure}[t]{\linewidth}
		\centering
		\includegraphics[width=0.8\linewidth]{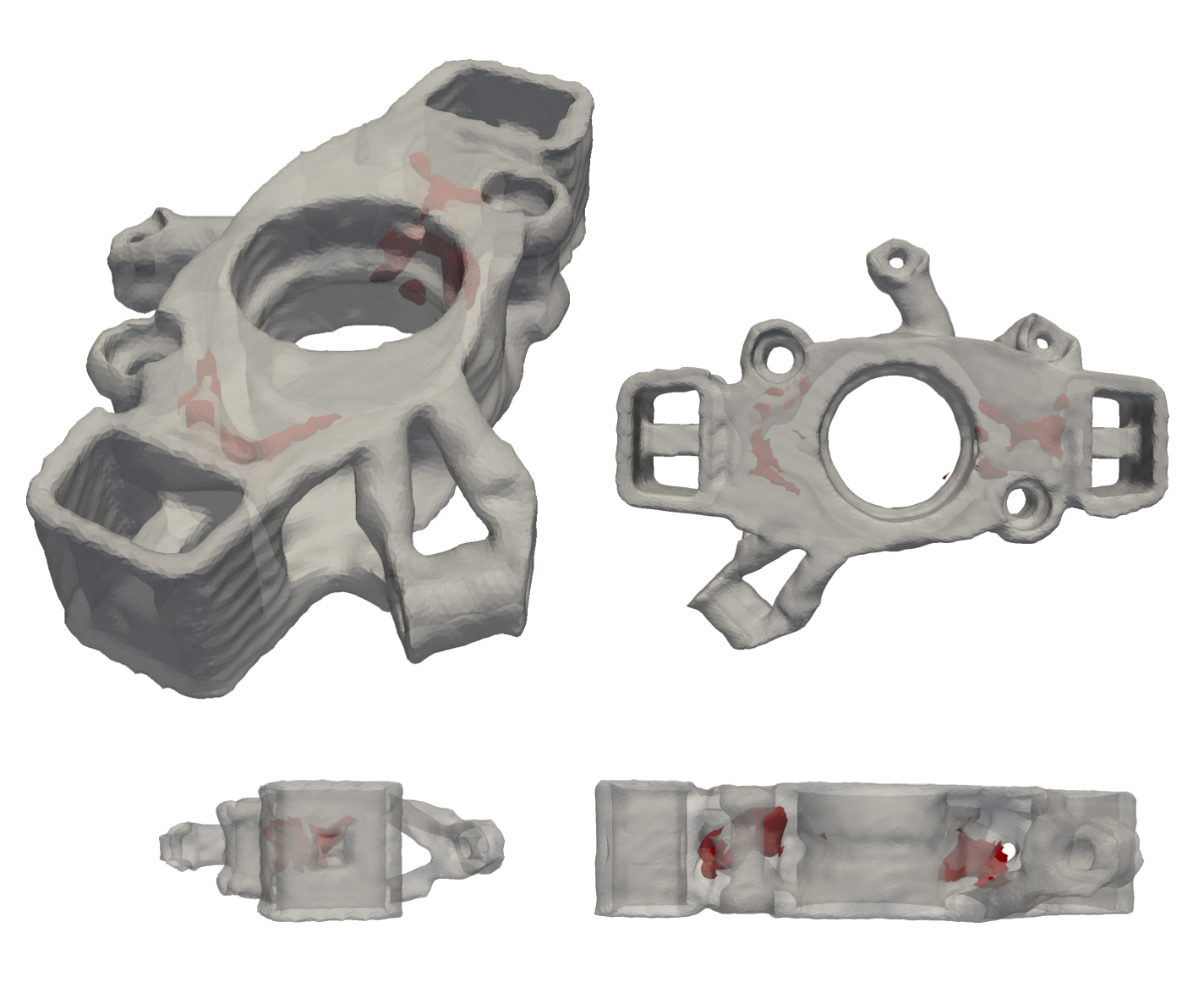}
		\caption{Without accessibility constraint.}
	\end{subfigure}%
	\\
	\begin{subfigure}[t]{\linewidth}
		\centering
		\includegraphics[width=0.8\linewidth]{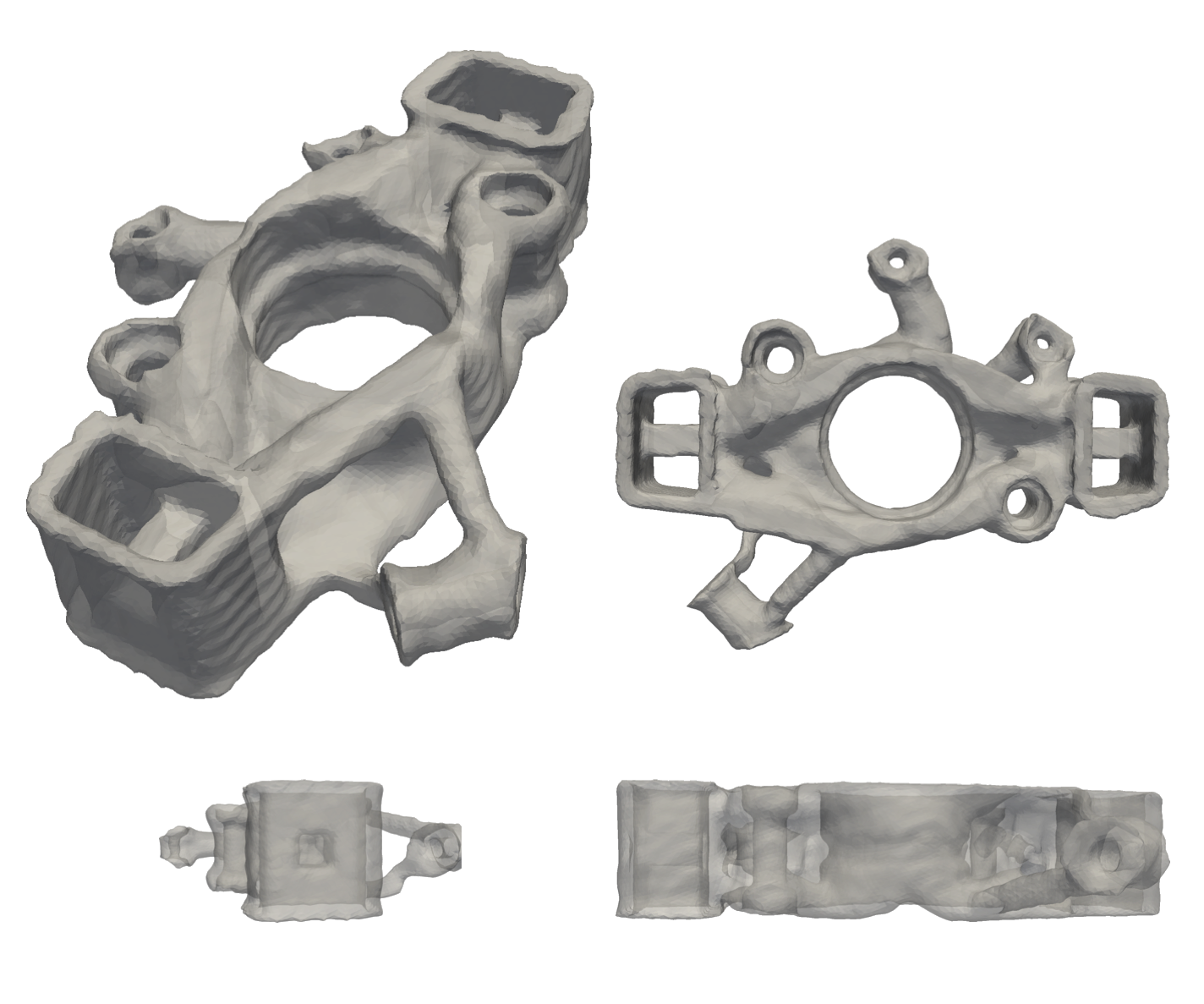}
		\caption{With accessibility constraint.}
	\end{subfigure}%
	\caption{Optimized upright bracket at 0.35 volume fraction.} \label{fig_upright}
\end{figure}

Figure \ref{fig_upright_supp} illustrates the additive and multi-axis machining setup where the build directions is along $\textbf{b} = \left\{0.25,0.25,0.85\right\}$, a VISE fixture holding the platform, and 3 different cutting tools approaching the part from different angles. Specifically, there is a total of 26 tool orientations with:
\begin{align} %
\Theta_1 = \{&R_{X}(0), R_{X}(\pi),\nonumber\\
&, R_{X}(\pm\pi/2),R_{X}(\pm\pi/4),\nonumber \\
&R_{X}(\pm 3\pi/4), R_{Z}(\pm\pi/2),\nonumber\\
&R_{Z}(\pm\pi/4),R_{Z}(\pm 3\pi/4)\} \nonumber  
\end{align}
\begin{align} %
\Theta_2 = \{&R_{X}(\pm\pi/2),R_{X}(-\pi/4),\nonumber \\
&R_{Z}(\pm 3\pi/4),-0.88\pi\} \nonumber  
\end{align}
\begin{align} %
\Theta_3 = \{&R_{X}(0), R_{X}(\pi),\nonumber\\
&R_{X}(\pm\pi/2),R_{Z}(\pm\pi/2),\nonumber 
\end{align}
The tools are initially oriented along +Y. Figure \ref{fig_upright} illustrates different views of the optimized upright without and with accessibility constraint. The unconstrained TO results in 7.14 $cm^3$, while the accessibility-constrained design exhibits an 18\% increase in compliance.

We have established that IMF ensures that a support removal plan exists for the optimized design shown in Fig. \ref{fig_upright}b for the machining setup shown in Figure \ref{fig_uprightLoading}. Although, the focus of this paper is not on generating a support removal plan, we propose a simple strategy to gradually remove support structures from the near-net shape assuming the part is only fixturable through the platform:

\begin{enumerate}
	\item Initialize the near-net shape and the machining setup 
	\item Starting from the top layer along build direction, find the tool and orientation that removes the maximum volume from that layer based on:
	\begin{enumerate}
		\item Compute IMF for each rotated tool via Algorithm \ref{alg_imf1}
		\item Take $\tau$ sub-levelset of IMF to find accessible regions
		\item Select orientated tool if accessible volume is the largest
	\end{enumerate}
	\item Update the near-net shape
	\item If there is no supports in this layer, go to next layer
\end{enumerate}

Note that the obstacle used to compute IMF for TO does not include support structure, since we are assuming that as long the tools do not collide with the part, platform, or fixture, there is a sequence by which the surrounding support structures can be removed to make a particular support region accessible. In other words, in TO formulation we are only interested in \textit{whether} the support can be removed not \textit{how}. On the other hand, to generate valid plans, obstacle for support removal planning is the union of the supports, part, platform, and fixturing device. Thus, performing task 2 for a single layer (along build direction) and small $\tau$ results in a very conservative plan that may be inefficient and cumbersome. The method can be relaxed by allowing removal from multiple layers and slightly larger $\tau$ to tolerate some level of collision with the support structures to allow volumetric removal in each step. To penalize collision with part or fixture, we assign higher values for these regions when we compute the convolution of \eq{eq_conv}. Here, we allow for removal of material from 10\% of total number of layers at each step with $\tau = 0.005$ and penalizing the non-support regions by a factor of 1000.  

\begin{figure} [ht!]
	\begin{subfigure}[t]{0.5\linewidth}
		\centering
		\includegraphics[width=0.8\linewidth]{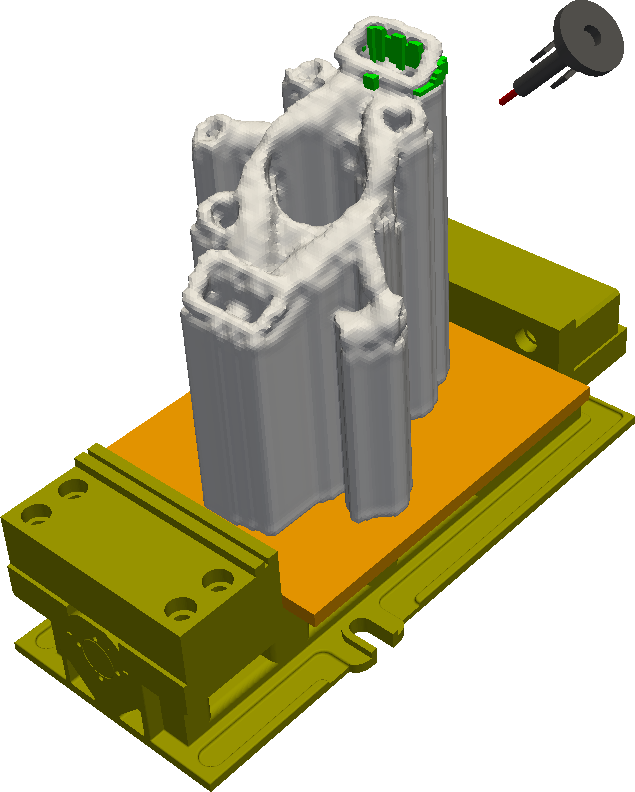}
		\caption{Tool 2 (0.14\%)}
	\end{subfigure}%
	\begin{subfigure}[t]{0.5\linewidth}
		\centering
		\includegraphics[width=0.8\linewidth]{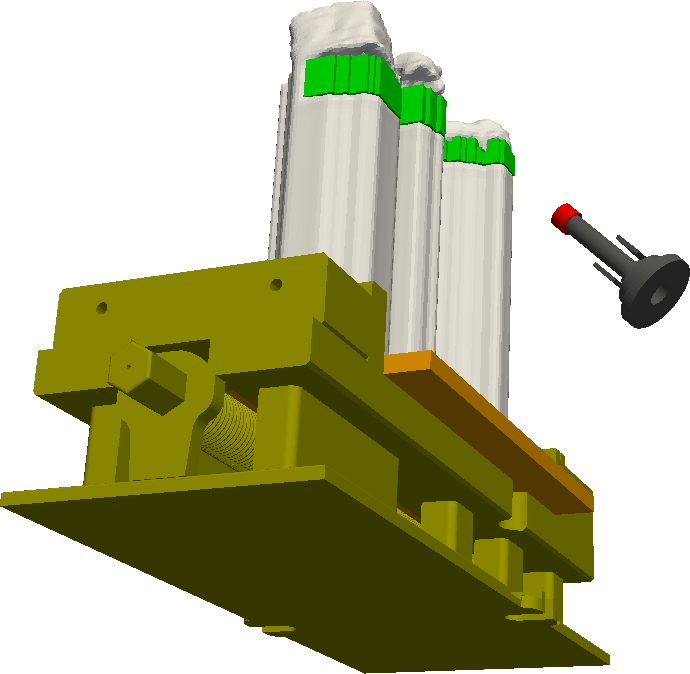}
		\caption{Tool 1 (0.79\%)}
	\end{subfigure}%
	\\
	\begin{subfigure}[t]{0.5\linewidth}
	\centering
	\includegraphics[width=0.8\linewidth]{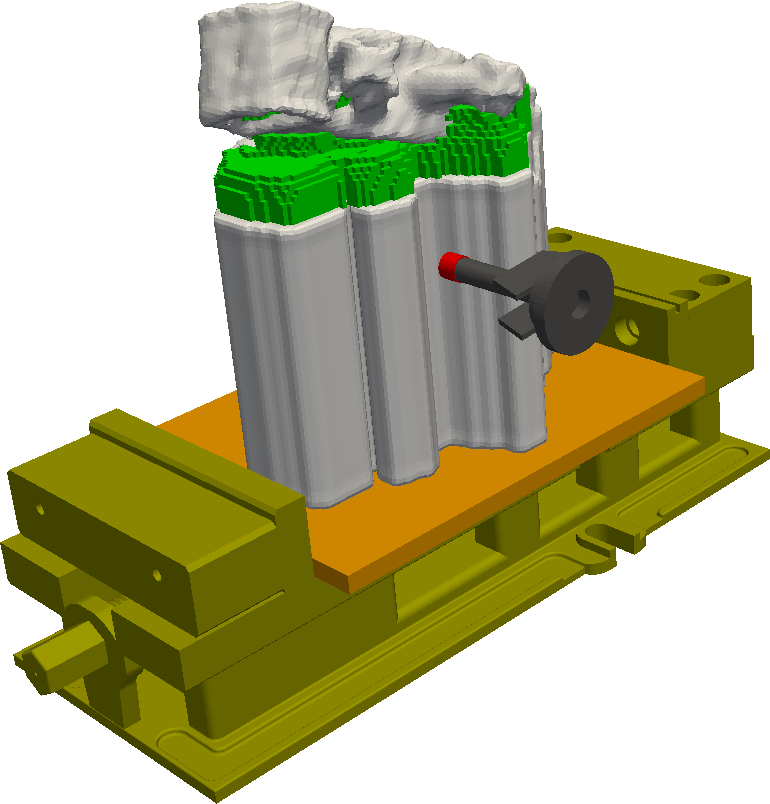}
	\caption{Tool 3 (2.97\%)}
	\end{subfigure}%
	\begin{subfigure}[t]{0.5\linewidth}
	\centering
	\includegraphics[width=0.8\linewidth]{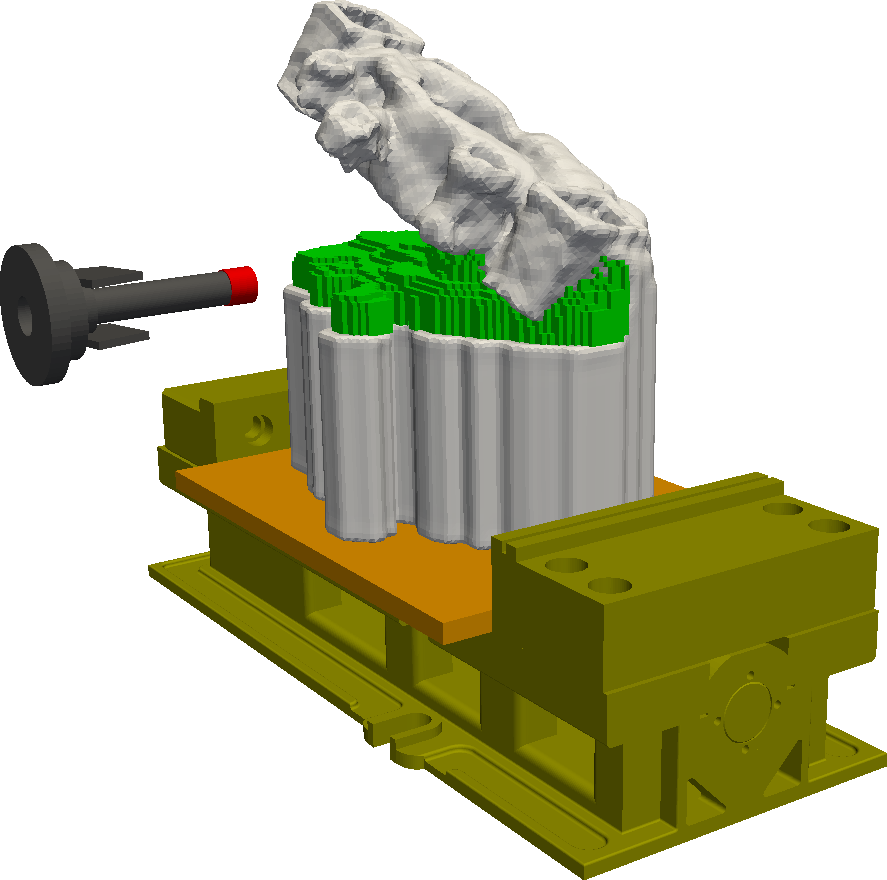}
	\caption{Tool 3 (5.82\%)}
	\end{subfigure}%
	\\
	\begin{subfigure}[t]{0.5\linewidth}
	\centering
	\includegraphics[width=0.8\linewidth]{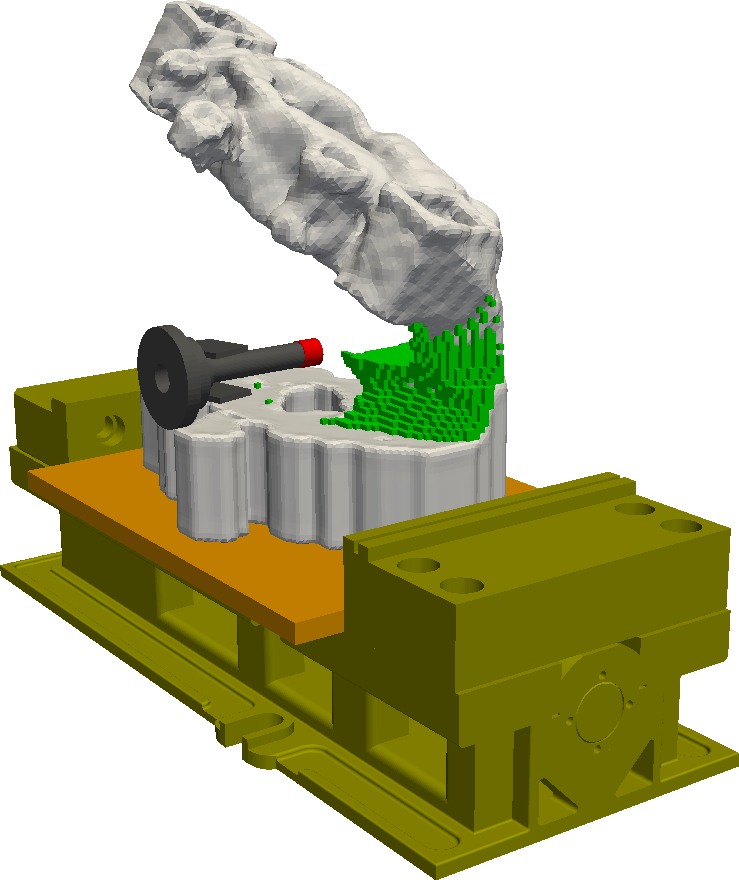}
	\caption{Tool 3 (1.05\%)}
	\end{subfigure}%
	\begin{subfigure}[t]{0.5\linewidth}
	\centering
	\includegraphics[width=0.8\linewidth]{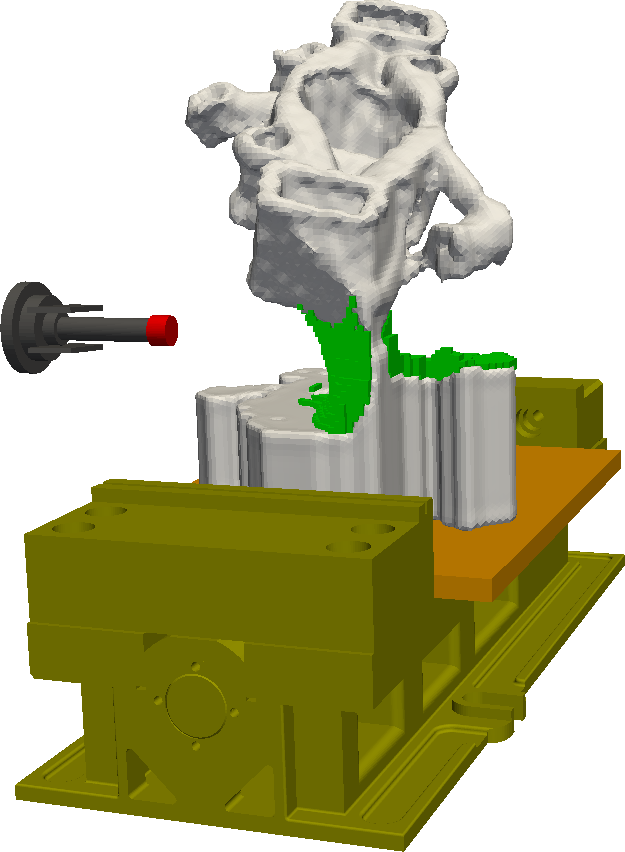}
	\caption{Tool 1 (1.76\%)}
	\end{subfigure}%
	\\
	\begin{subfigure}[t]{0.5\linewidth}
	\centering
	\includegraphics[width=0.8\linewidth]{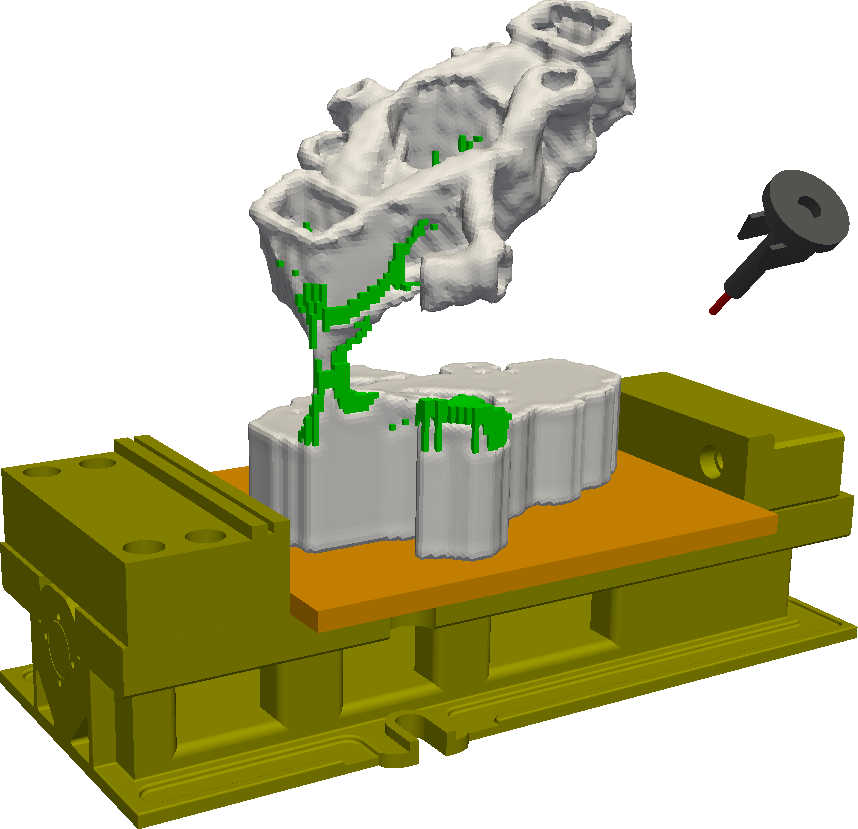}
	\caption{Tool 2 (0.70\%)}
	\end{subfigure}%
	\begin{subfigure}[t]{0.5\linewidth}
	\centering
	\includegraphics[width=0.8\linewidth]{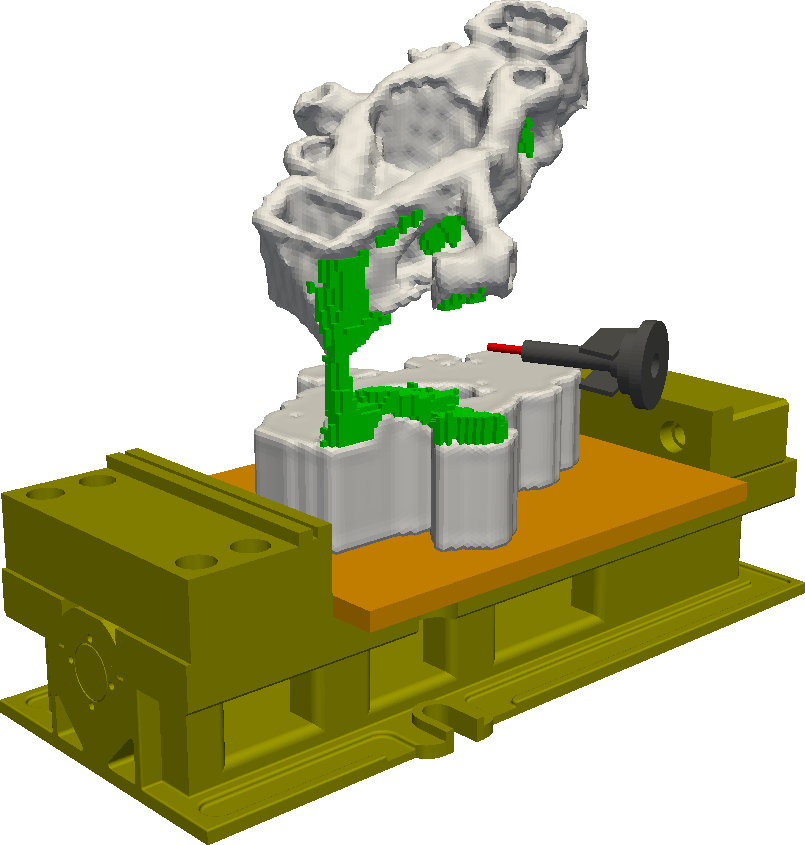}
	\caption{Tool 2 (1.23\%)}
	\end{subfigure}%
	\caption{Sampled steps of support removal plan where supports are gradually removed (green) from top to bottom of the current near-net shape (gray) until the part is detached from the platform. In this case, overall 63\% of the supports need to be machined. } \label{fig_uprightplan}
\end{figure}

Figure \ref{fig_uprightplan} illustrates 8 sampled intermediate steps is support removal planning, where different tools are used to machine as much support structures as possible while ensuring that the part remains connected to the platform until all supports that are in contact with the part are removed. In this case, overall 63\% of the supports need to be machined. 

\section{Conclusion}
We presented a general approach to ensure removability of support structures upfront at the early stages of design for a hybrid AM-then-SM process. In particular, we extend the concept of `inaccessibility measure field' (IMF) introduced in \cite{mirzendehdel2020topology} to near-net shapes produced by many metal additive processes. IMF is a continuous field defined using well-established mathematical concepts in spatial planning in terms of convolutions in configuration space. 

We extend the density-based TO formulation to incorporate accessibility of supports given a multi-axis machining setup comprised of a collection of realistic tool assemblies and fixturing devices. We define a normalized support accessibility filter that is augmented to the performance sensitivity field by an adaptive weight factor $w_{acc}$. To improve convergence, for the first few iterations (say 20), the accessibility constraint is ignored ($w_{acc} = 0$). Once an initial approximation of the design is obtained, $w_{acc}$ is gradually increased up to an upper-bound (say 0.5). Our preliminary experiments showed that a layer-wise penalization of the support inaccessibility measure field helps the optimizer to converge to a better design. Further, since the constraint is imposed as a soft constraint, in some pathological cases a small amount of inaccessible supports remain towards the end of optimization. In this case we continue optimization while also imposing a filter directly on density values. In our experiments, this helps TO to converge to a valid solution within a few iterations. The effectiveness of our proposed method is demonstrated through benchmark and realistic examples in 2D and 3D.

In this work, the part remains fixturable throughout the optimization process since we assumed that it is sufficient for the fixturing device to only the platform. This may be limiting in support removal planning, where in many cases we need to hold the workpiece itself or design a soft-jaw to ensure support removal without damaging the part.  
Future work will focus on incorporating a general fixturability constraint in TO. 


\bibliographystyle{elsarticle-num} 
\bibliography{topOptAccSupps}

\end{document}